\documentclass[]{interact}
\graphicspath{{Figures/}}
\usepackage{epstopdf}
\usepackage[caption=false]{subfig}
\usepackage[authoryear,sort]{natbib}
\bibpunct[, ]{(}{)}{;}{a}{,}{,}

\theoremstyle{plain}

\theoremstyle{definition}

\theoremstyle{remark}

\usepackage[bookmarks=true,colorlinks,linkcolor=blue,anchorcolor=blue,citecolor=blue,unicode,urlcolor=blue]{hyperref}
\usepackage{multirow}
\usepackage{enumerate}
\usepackage{booktabs}
\usepackage{lscape,pdflscape}
\usepackage{afterpage}
\usepackage{graphicx,color}

\begin{document}

\articletype{RESEARCH ARTICLE}

\title{Quantifying the dynamic structural resilience of international staple food trade networks: An entropy-based approach}
		
\author{
\name{Si-Yao Wei\textsuperscript{a,b} and Wei-Xing Zhou\textsuperscript{a,b,c}\thanks{CONTACT Wei-Xing Zhou. Email: wxzhou@ecust.edu.cn}}
\affil{\textsuperscript{a}School of Business, East China University of Science and Technology, Shanghai, China; \textsuperscript{b}Research Center for Econophysics, East China University of Science and Technology, Shanghai, China; \textsuperscript{c}School of Mathematics, East China University of Science and Technology, Shanghai, China.}}

\maketitle

\begin{abstract}
Establishing a resilient food trade system is an international consensus on safeguarding food security amid growing disruptions. However, a unified resilience framework has yet to be established, leading to the proliferation of diverse measures. Here, we conceptualize resilience as a trade-off between efficiency and redundancy and employ an entropy-based approach to quantify the dynamic structural resilience of international trade networks for maize, rice, soybean, and wheat from 1986 to 2022. Using index decomposition analysis, we also investigate the relative contributions of internal components to resilience dynamics. Within this framework, despite heterogeneity across different food commodities, we find that current trade networks are relatively redundant, with improvements in efficiency being the dominant driver of changes in resilience. In addition, we reveal a historically pronounced impact of flow concentrations on resilience, while trade interactions have become increasingly important in recent years. Following the leave-one-out approach, we furthermore identify critical economies and trade relationships that disproportionately affect the overall resilience, some of which are less well-focused in previous studies. Moreover, we highlight that overconcentration of flows along core trade relationships may undermine both efficiency and resilience, whereas peripheral trade networks may play strategic alternative roles in sustaining resilience, underscoring the importance of concentrating on developing economies and promoting broader trade links. These findings not only provide new insights for assessing the resilience of international food trade systems but also propose directions for strengthening resilience through both regional cooperation and more inclusive trade relations.
\end{abstract}

\begin{keywords} 
International food trade network; Food security; Network resilience; Leave-one-out approach; Index decomposition analysis
\end{keywords}

\clearpage


\section{Introduction}
\label{Sec_Introduction}

Food security, an increasingly prominent global concern, was traditionally conceptualized as comprising three core pillars: availability, access, and utilization \citep{FJ-Barrett-2010-Science}. Recently, this framework has been extended to incorporate three additional dimensions\textemdash stability, agency, and sustainability\textemdash thus forming a comprehensive six-dimensional paradigm \citep{FJ-Clapp-Moseley-Burlingame-Termine-2022-FoodPol}. Historically, food insecurity has been driven primarily by shortages in food availability, but in modern times, global famine and destitution are primarily caused by limited access to food \citep{FJ-Suweis-Carr-Maritan-Rinaldo-DOdorico-2015-PNAS}. Pandemics, extreme climate changes, pests, and geopolitical conflicts significantly destabilize food availability and access \citep{FJ-Naqvi-Gaupp-HochrainerStigler-2020-ORSpectrum,FJ-Nicholson-Emery-Niles-2021-NatCommun}. Stabilizing these two pillars is therefore a key component of contemporary food security strategies \citep{FJ-Barrett-2010-Science}.

Within the framework of food security, international food trade contributes to both food availability and accessibility \citep{FJ-Traverso-Schiavo-2020-WorldDev}, although its role in safeguarding food security remains a subject of debate \citep{FJ-Farsund-Daugbjerg-Langhelle-2015-FoodSecur,FJ-Grassia-Mangioni-Schiavo-Traverso-2022-SciRep}. Some scholars emphasize the positive contribution of food trade to food security \citep{FJ-Dorosh-Dradri-Haggblade-2009-FoodPol,FJ-Grassia-Mangioni-Schiavo-Traverso-2022-SciRep}. For example, \cite{FJ-Dithmer-Abdulai-2017-FoodPol} argue that food trade can help mitigate domestic excess demand or supply, thereby providing countries with more options to enhance food availability than under a self-sufficiency policy, and contributing to greater stability in national food supply and prices. Similarly, \cite{FJ-Traverso-Schiavo-2020-WorldDev} find that participation in international food trade positively affects aggregate food availability and access in low-income countries, from a perspective of macronutrient flows. In contrast, other scholars highlight its potential downsides \citep{FJ-Headey-2011-FoodPol,FJ-Odey-Adelodun-Lee-Adeyemi-Choi-2023-JEM}. For instance, \cite{FJ-Bouet-Bureau-Decreux-Jean-2005-WorldEcon} argue that the benefits of trade may be overestimated, particularly for developing countries lacking a comparative advantage in agricultural production. Based on evidence from developing economies, \cite{FJ-Mary-2019-FoodSecur} suggests that pursuing food self-sufficiency is more beneficial in the short term, even if such policies conflict with World Trade Organization regulations and the prevailing global trade agenda.

Although the role of international trade in food security remains contested, there is broad consensus on the necessity of establishing a resilient international food trade system \citep{FJ-Tu-Suweis-DOdorico-2019-NatSustain, FAO-2021-State, FJ-Engemann-Jafari-Heckelei-2023-JAgricEcon}. Achieving this goal, however, requires addressing two foundational questions: (1) what constitutes resilience in the context of the international food trade system? and (2) how can it be effectively measured? 

Resilience is a fundamental property of complex systems \citep{FJ-Liu-Li-Ma-Szymanski-Stanley-Gao-2022-PhysRep} and its concept originates in ecology, where \cite{FJ-Holling-1973-AnnuRevEcolSystemat} defines it as a system's capacity to withstand disturbances without collapsing or transitioning to a fundamentally altered state, and to recover its essential functions afterward. Building on this concept, various disciplines\textemdash including engineering, psychology, management, and economics\textemdash have developed discipline-specific definitions of resilience \citep{FJ-Youssef-Luthans-2007-JManag, FJ-Francis-Bekera-2014-ReliabEngSystSaf, FJ-Brunnermeier-2024-JFinanc,FJ-Padovano-Ivanov-2025-IntJProdRes,FJ-Gunasekaran-Subramanian-Rahman-2015-IntJProdRes,FJ-Ivanov-2024-IntJProdRes}. However, a unified theory and framework have yet to be established \citep{FJ-Liu-Li-Ma-Szymanski-Stanley-Gao-2022-PhysRep}, which has led to the proliferation of diverse resilience measures \citep{FJ-Pimm-Donohue-Montoya-Loreau-2019-NatSustain}.

Many scholars emphasize resilience as the capacity of systems to adapt and recover under disruptions \citep{FJ-Haimes-2009-RiskAnal}, which is commonly assessed using performance-based approaches. These approaches focus primarily on a system's performance before and after disruptions, regardless of its structural features \citep{FJ-Hosseini-Barker-RamirezMarquez-2016-ReliabEngSystSaf,FJ-Qi-Mei-2024-JKingSaudUnivComputInfSci}. Among them, the {\it Resilience Triangle} is a widely used framework \citep{FJ-Bruneau-Chang-Eguchi-Lee-ORourke-Reinhorn-Shinozuka-Tierney-Wallace-VonWinterfeldt-2003-EarthqSpectra}, which measures resilience as the level of performance retained during a disturbance and visualizes the magnitude, duration, and recovery of performance loss \citep{FJ-Cheng-Elsayed-Huang-2022-IntJProdRes}. Within this framework, for example, studies in engineering often assess resilience by measuring the cumulative loss in infrastructure quality over time \citep{FJ-Fang-Chu-Fu-Fang-2022-TranspResPartD,FJ-Chen-Ma-Chen-Yang-2024-TranspResPartD, FJ-Sahebjamnia-Torabi-Mansouri-2015-EurJOperRes}. In economics, resilience is often understood as an economic system's dynamic response to external shocks, and can be thus measured through response functions \citep{FJ-Klimek-Poledna-Thurner-2019-NatCommun, FJ-Brunnermeier-2024-JFinanc}. In ecology, resilience is typically quantified in terms of recovery rate or the time required to adapt to changing disturbance regimes \citep{FJ-Seidl-Spies-Peterson-Stephens-Hicke-2016-JApplEcol}. These studies emphasize the system's speed of recovery and the magnitude of loss, offering valuable insights into resilience measurement. 

In addition to performance-based approaches, scholars also focus on how a system's structural properties shape its resilience, where structure-based approaches are widely used \citep{FJ-Hosseini-Barker-RamirezMarquez-2016-ReliabEngSystSaf}. Indeed, the choice of approach often reflects which principles of resilience are emphasized, yet no consensus emerges. For example, some studies regard diversification as the most fundamental principle of resilience \citep{FJ-Hertel-Elouafi-Tanticharoen-Ewert-2021-NatFood} due to their strong correlations \citep{FJ-Essuman-OwusuYirenkyi-Afloe-Donbesuur-2023-JIntManag,FJ-Stevens-Teal-2024-AmJAgrEcon,FJ-Habibi-Chakrabortty-Abbasi-Ho-2025-IntJProdRes,FJ-Arndt-Helming-2025-AgricEcosystEnviron}, as higher diversification implies that a system is less likely to collapse in the face of disturbances \citep{FJ-Nicholson-Emery-Niles-2021-NatCommun}. Redundancy has also been mentioned as a basic principle of resilience, especially in supply chains, where sufficient back-up facilities enhance supply chain resilience during disruptions \citep{FJ-Kamalahmadi-Parast-2016-IntJProdEcon,FJ-Tan-Zhang-Cai-2019-IntJProdRes,FJ-Ivanov-2024-Omega-IntJManageSci}. In addition, the existing literature also highlights system cohesion, clustering, and efficiency as important elements of resilience \citep{FJ-Bhamra-Dani-Burnard-2011-IntJProdRes,FJ-Kahiluoto-Makinen-Kaseva-2020-IntJOperProdManage,FJ-Yu-Ma-Wang-2024-IntJProdEcon}, which further complement the principles of resilience.

Furthermore, given that most real-world systems are structured as networks with complex topologies \citep{FJ-Barabasi-Albert-1999-Science}, a substantial body of literature assesses system resilience by analyzing the evolution of a network's topological metrics, as a key stream of structure-based approaches. For example, resilience can be assessed using the inverse of the harmonic mean of the shortest path lengths between nodes, referred to as global efficiency (which differs from the notion of efficiency used in our study) \citep{FJ-Latora-Marchiori-2001-PhysRevLett,FJ-OseiAsamoah-Lownes-2014-TranspResRec,FJ-Testa-Furtado-Alipour-2015-TranspResRec,FJ-Xie-Wei-Zhou-2021-JStatMech,FJ-Wei-Xie-Zhou-2022-JComplexNetw,FJ-Chen-Ma-Chen-Yang-2024-TranspResPartD}. Moreover, \cite{FJ-Dixit-Verma-Tiwari-2020-IntJProdEcon} measure supply chain resilience by integrating four topological indicators, including density, centrality, size, and connectivity, which is also applied to assess resilience in crude oil and gas trade networks \citep{FJ-Mou-Sun-Yang-Wang-Zheng-Chen-Zhang-2020-IEEEAccess,FJ-Sun-Wei-Jin-Song-Li-2023-GlobNetw}. Notably, within a framework proposed by \cite{FJ-Gao-Barzel-Barabasi-2016-Nature}, existing literature can infer whether a system is resilient based on dynamic equations \citep{FJ-Xu-Si-Duan-Lv-Xie-2019-PhysicaA,FJ-Zhao-Ren-Zhao-Weng-2024-HumSocSciCommun,FJ-Tu-Suweis-DOdorico-2019-NatSustain}, which advances structure-based approaches by incorporating system dynamics. 

Now, let us concentrate our attention on the field of food systems. Within this, resilience is generally defined as the ability to maintain a stable food supply\textemdash in both quantity and diversity\textemdash even under disruptions \citep{FJ-Tendall-Joerin-Kopainsky-Edwards-Shreck-Le-Krutli-Grant-Six-2015-GlobFoodSecur}, which has attracted considerable scholarly attention \citep{FJ-Zurek-Ingram-Bellamy-Goold-Lyon-Alexander-Barnes-Bebber-Breeze-Bruce-Collins-Davies-Doherty-Ensor-Franco-Gatto-Hess-Lamprinopoulou-Liu-Merkle-Norton-Oliver-Ollerton-Potts-Reed-Sutcliffe-Withers-2022-AnnuRevEnvironResour,FJ-Seekell-Carr-DellAngelo-DOdorico-Fader-Gephart-Kummu-Magliocca-Porkka-Puma-Ratajczak-Rulli-Suweis-Tavoni-2017-EnvironResLett}. Analogously, in the field of food trade systems, resilience is generally defined as the ability to maintain a stable availability of, and access to, food supply via international trade\textemdash in both quantity and diversity\textemdash even under disruptions \citep{FJ-Mena-Karatzas-Hansen-2022-JBusRes,FJ-Jafari-Engemann-Zimmermann-2024-QOpen}. With this definition, both performance-based and structure-based approaches have been employed to measure the resilience of food trade systems. For example, under simulated shock scenarios, \cite{FJ-Fair-Bauch-Anand-2017-SciRep} and \cite{FJ-GutierrezMoya-AdensoDiaz-Lozano-2021-FoodSecur} assess the resilience of wheat trade networks, while \cite{FJ-Gephart-Rovenskaya-Dieckmann-Pace-Brannstrom-2016-EnvironResLett} develop a forward shock-propagation model to evaluate the vulnerability of the global seafood trade network. Furthermore, \cite{FJ-Laber-Klimek-Bruckner-Yang-Thurner-2023-NatFood} and \cite{FJ-Baum-Laber-Bruckner-Yang-Thurner-Klimek-2025-arXiv} employ multilayer network approaches to examine how shocks propagate through intertwined production and trade channels, while \cite{FJ-Kuhla-Kubiczek-Otto-2025-EcolEcon} use an agent-based model to reproduce crisis dynamics in agricultural markets. Employing topological indicators, \cite{FJ-Lin-Dang-Konar-2014-EnvironSciTechnol} identify structural vulnerabilities in U.S. food flow networks, and \cite{FJ-Ji-Zhong-Nzudie-Wang-Tian-2024-JCleanProd} analyze the vulnerability of global food trade networks. Moreover, \cite{FJ-Karakoc-Konar-2021-EnvironResLett} develop a resilience framework based on eigenvalues and nodes' dependence on major exporters. Similar approaches are also applied in other studies \citep{FJ-Wang-Dai-2021-Foods,FJ-Grassia-Mangioni-Schiavo-Traverso-2022-SciRep,FJ-Xu-Niu-Li-Wang-2024-Sustainability}.

Despite extensive academic discussions on the resilience of food trade systems, existing related studies remain incomplete and in need of further complement. First, the principles that constitute resilience remain unsettled, leading to ongoing debates about how it should be measured \citep{FJ-Kahiluoto-Makinen-Kaseva-2020-IntJOperProdManage}. Although diversification and redundancy are often regarded as the most fundamental principles of resilience, research on supply chains suggests that their single-minded pursuit may erode profits without improving resilience \citep{FJ-Pettit-Croxton-Fiksel-2019-JBusLogist}. Instead, an emerging consensus among both scholars and practitioners emphasizes that enhancing resilience requires achieving an appropriate trade-off among multiple principles \citep{FJ-Brede-deVries-2009-PhysLettA,FJ-Kahiluoto-Makinen-Kaseva-2020-IntJOperProdManage,FJ-Yang-Wu-Sun-Zhang-2024-JRiskRes}, a perspective that warrants further empirical substantiation \citep{FJ-Kamalahmadi-Parast-2016-IntJProdEcon}.

Second, existing approaches remain inadequate. On the one hand, performance-based approaches emphasize system performance after disturbances while neglecting baseline conditions in the absence of shocks, which obscures the fact that resilience is a fundamental system property \citep{FJ-Liu-Li-Ma-Szymanski-Stanley-Gao-2022-PhysRep}. In addition, determining shock duration and recovery thresholds remains challenging and often unrealistic \citep{FJ-Hosseini-Barker-RamirezMarquez-2016-ReliabEngSystSaf}. On the other hand, structure-based measures such as topology-based measures often overlook network dynamics and fail to capture the temporal evolution of resilience \citep{FJ-Qi-Mei-2024-JKingSaudUnivComputInfSci}. There are, of course, notable exceptions. For example, \cite{FJ-Wei-Xiao-Li-Huang-Liu-Xue-2025-ResourConservRecycl} integrate the {\it Resilience Triangle} with dynamic equations in a unified framework, offering new methodological insights. Yet, resilience inference within the framework of \cite{FJ-Gao-Barzel-Barabasi-2016-Nature} relies on the strong assumption that the degrees of interconnected nodes are largely independent \citep{FJ-Liu-Xu-Gao-Wang-Li-Gao-2024-NatCommun}. Such an assumption contrasts with the high assortativity observed in international trade networks \citep{FJ-Mou-Fang-Yang-Zhang-2020-IEEEAccess}, which may limit the applicability of this framework in the context of global food trade. Consequently, there is a pressing need for resilience measures that not only reflect the dynamic evolution of network structures but also capture the fundamental principles underpinning resilience \citep{FJ-Li-Wang-Kharrazi-Fath-Liu-Liu-Xiao-Lai-2024-FoodSecur}.

Third, existing research needs to strengthen its focus on the internal mechanisms driving resilience and on the roles of peripheral economies in maintaining overall resilience. On the one hand, many studies primarily examine correlations between food trade system resilience and determinants such as topological indicators and population size \citep{FJ-Ji-Zhong-Nzudie-Wang-Tian-2024-JCleanProd,FJ-GutierrezMoya-AdensoDiaz-Lozano-2021-FoodSecur}, while few investigate the internal mechanisms that drive resilience. On the other hand, core trade economies have been the main focus through various centrality measures \citep{FJ-Grassia-Mangioni-Schiavo-Traverso-2022-SciRep,FJ-GutierrezMoya-AdensoDiaz-Lozano-2021-FoodSecur,FJ-Zhang-Zhou-2023-ChaosSolitonFract}, yet because many such measures emphasize developed economies with numerous partners, they still potentially overlook economies that have peripheral positions in the network but contribute to maintaining the overall resilience. Moreover, link-level disruptions occur more frequently in real-world trade systems, yet existing research predominantly examines the significance of nodes, leaving the importance of trade links relatively less explored \citep{FJ-Qian-Li-Zhang-Ma-Lu-2017-SciRep,FJ-Guo-Chen-Fan-Zeng-Pu-Qing-2025-ExpertSystAppl}.

To address these gaps and concerns, the following research questions frame our work:

\begin{enumerate}[RQ1:]

\item How can the resilience of the international food trade system be effectively measured?

\item How do internal components contribute to changes in resilience of the international food trade system?

\item How can critical economies and trade relationships in the international food trade system be identified from a resilience perspective?

\end{enumerate}

The subsequent sections of this paper are structured in the following manner. The methodological framework is presented in Section~\ref{Sec_Methodology}. Section~\ref{Sec_Evolution} describes the international food trade network and its resilience. Then, Section~\ref{Sec_Correlation} and Section~\ref{Sec_Identify} present and analyze the main findings. Section~\ref{Sec_Discussion} discusses the contributions, implications, limitations, and future outlook. Conclusion is finally presented in Section~\ref{Sec_Conclusion}.

\section{Methodology}
\label{Sec_Methodology}

\subsection{Topological indicators of the international food trade system}

The international food trade system is usually regarded as a network that exhibits small-world (i.e., high clustering coefficients coupled with short average path lengths) and scale-free (i.e., power-law degree distributions) topological indicators in economies' connectivity \citep{FJ-Fagiolo-Reyes-Schiavo-2009-PhysRevE}. A food trade network is a weighted digraph $G=\langle\mathcal{V},\mathcal{E}\rangle$, where $\mathcal{V}$ represents the vertex set with $N_\mathcal{V}$ economies and $\mathcal{E}$ represents the link set with $N_\mathcal{E}$ trade relationships. $f_{ij}$ denotes the volume of trade flow that economy $i$ exports to economy $j$. There are several commonly used topological metrics for characterizing network structure, which are briefly described below.

\begin{enumerate}[(1)]
\item The in-degree $k_i^{\mathrm{in}}$ and out-degree $k_i^{\mathrm{out}}$ of economy $i$ are defined respectively as the numbers of economies from which it imports and to which it exports.

\item The in-strength $s_i^{\mathrm{in}}=\sum\limits_{i=1}^{N_\mathcal{V}}f_{ij}$ and out-strength $s_i^{\mathrm{out}}=\sum\limits_{j=1}^{N_\mathcal{V}}f_{ij}$ of economy $i$ are defined as the total volumes of imports and exports, respectively.

\item The average degree and average strength are respectively calculated as: $\langle k\rangle=\frac{1}{N_\mathcal{V}}\sum\limits_{i=1}^{N_\mathcal{V}}k_i$ and $\langle s\rangle=\frac{1}{N_\mathcal{V}}\sum\limits_{i=1}^{N_\mathcal{V}}s_i$, where $k_i=k_i^{\mathrm{in}}+k_i^{\mathrm{out}}$ and $s_i=s_i^{\mathrm{in}}+s_i^{\mathrm{out}}$.

\item The average clustering coefficient represents the average level of interconnectedness among the trade partners of economies, defined as $\langle C\rangle=\frac{1}{N_\mathcal{V}}\sum\limits_{i=1}^{N_\mathcal{V}}C_i$, where the local clustering coefficient of $i$ is $C_i=\frac{T_i}{k_i^{\mathrm{in}}(k_i^{\mathrm{in}}-1)+k_i^{\mathrm{out}}(k_i^{\mathrm{out}}-1)}$ and $T_i$ represents the number of actual connections among the trade partners of economy $i$.

\item The density represents the level of trade closeness among economies, defined as: $\rho=\frac{N_\mathcal{E}}{N_\mathcal{E}^{\max}}=\frac{N_\mathcal{E}}{N_\mathcal{V}(N_\mathcal{V}-1)}$, where $N_\mathcal{E}^{\max}$ represents the maximum number of links possible in the network.

\item The average shortest path length represents the average level of separation among economies, defined as $L=\frac{1}{N_\mathcal{V}(N_\mathcal{V}-1)}\sum\limits_{i\neq j\in\mathcal{V}} d_{ij}$, where $d_{ij}$ represents the length of the shortest path connecting economy $i$ to economy $j$.

\item The diameter represents the length of the longest $d_{ij}$, defined as $D=\max\limits_{i\neq j\in\mathcal{V}} d_{ij}$.

\item Modularity represents the effectiveness of partitioning a network into communities, defined as $Q = \frac{1}{N_\mathcal{V}\langle s\rangle} \sum\limits_{i=1}^{N_\mathcal{V}}\sum\limits_{j=1}^{N_\mathcal{V}}\left(f_{ij}-\frac{s_is_j}{N_\mathcal{V}\langle s\rangle}\right)\delta(c_i,c_j)$, where $\delta(c_i,c_j)=1$ if economies $i$ and $j$ belong to the same community and $\delta(c_i,c_j)=0$ otherwise. The Louvain algorithm, recognized as one of the fastest and best performing algorithms in comparative analysis \citep{FJ-Blondel-Guillaume-Lambiotte-Lefebvre-2008-JStatMech,FJ-Lancichinetti-Fortunato-2009-PhysRevE}, is employed for community detection in this paper.
\end{enumerate}

\subsection{Efficiency and redundancy of the international food trade system}

To address the above methodological gaps, information theory has been introduced into resilience research. In information theory, entropy measures a system's uncertainty (or diversity). A higher entropy reflects greater unpredictability and diversity, indicating enhanced adaptability to changing conditions and an improved capacity of the system to absorb shocks \citep{FJ-Khakifirooz-Fathi-Dolgui-Pardalos-2025-IntJProdRes,FJ-Reggiani-2022-NetwSpatEcon}. Accordingly, we can adopt joint entropy to characterize a system's overall structural properties \citep{FJ-Rutledge-Basore-Mulholland-1976-JTheorBiol,FJ-Ulanowicz-1979-Oecologia}, which is also called as ``capacity'' for system development \citep{FJ-Ulanowicz-Norden-1990-IntJSystSci}. Given that joint entropy is equal to the sum of mutual information and conditional entropy (e.g., $H(X,Y)=I(X,Y)+H(Y|X)+H(X|Y)$), we further introduce two underlying quantities, efficiency (e.g., $I(X,Y)$) and redundancy (e.g., $H(Y|X)+H(X|Y)$), which demonstrate opposing properties of a system. 

Efficiency embodies the flow articulation or constraints within networked configurations, which tends to increase due to preferential interactions between economies \citep{FJ-Kharrazi-Rovenskaya-Fath-2017-PloSOne}. As economies become more interdependent, trade flows tend to become more concentrated and specialized, thereby enhancing overall system efficiency. In information theory, a higher mutual information indicates a increased reduction in uncertainty when some information is known and a stronger statistical inter-dependence between variables. Denoting $s=\sum\limits_{i=1}^{N_\mathcal{V}}s_i^{\mathrm{out}}=\sum\limits_{j=1}^{N_\mathcal{V}}s_j^{\mathrm{in}}=\sum\limits_{i=1}^{N_\mathcal{V}}\sum\limits_{j=1}^{N_\mathcal{V}}f_{ij}$, the efficiency $e$ can be expressed as
\begin{equation}
    e=\sum\limits_{i=1}^{N_\mathcal{V}}\sum\limits_{j=1}^{N_\mathcal{V}} p_{ij}\ln\frac{p_{ij}}{p_{i}^{\mathrm{out}}p_{j}^{\mathrm{in}}}
    =\sum_{i=1}^{N_\mathcal{V}}\sum_{j=1}^{N_\mathcal{V}}\frac{f_{ij}}{s} \ln\frac{f_{ij}s}{s_i^{\mathrm{out}}s_j^{\mathrm{in}}},
    \label{eq_efficiency}
\end{equation}
which is also called as the average mutual information \citep{FJ-Rutledge-Basore-Mulholland-1976-JTheorBiol,FJ-Ulanowicz-Norden-1990-IntJSystSci} and the larger value occurs when trade flows are highly predictable due to the stronger inter-dependence (i.e., $p_{ij}\gg p_i^{\mathrm{out}}p_j^{\mathrm{in}}$ for links with larger volume), where 
\begin{equation}\label{eq_p_ij}
    p_{ij}=\frac{f_{ij}}{s}
\end{equation}
denotes the proportion of the volume $f_{ij}$ of trade flow from economy $i$ to $j$ to the total trade volume $s$, and $p_{i}^{\mathrm{out}}$ and $p_{j}^{\mathrm{in}}$ respectively denote the proportion of export volume from economy $i$ to the others:
\begin{equation}
    p_{i}^{\mathrm{out}}=\frac{s_i^{\mathrm{out}}}{s},
\end{equation}
and that of import volume from the others to economy $j$:
\begin{equation}
    p_{j}^{\mathrm{in}}=\frac{s_j^{\mathrm{in}}}{s}.
\end{equation}

Redundancy embodies the diversity of food trade pathways, which is critical for the system's capacity adapting to changing environmental conditions arising from shocks or disturbances \citep{FJ-Luo-Yu-Kharrazi-Fath-Matsubae-Liang-Chen-Zhu-Ma-Hu-2024-NatFood}. A greater number of pathways in food trade allows economies to have greater freedom in the trade market and enhances the system's robustness in coping with risks. Note that our perspective departs from studies that focus primarily on overall uncertainty and instead emphasizes the trade-off between efficiency and redundancy. In this context, redundancy is conceptualized as the residual uncertainty conditional on partial information, which corresponds to the definition of conditional entropy in information theory. A higher conditional entropy indicates a greater availability of alternative pathways in food trade and a consequently higher system redundancy. Accordingly, the redundancy $r$ can be expressed as \citep{FJ-Rutledge-Basore-Mulholland-1976-JTheorBiol,FJ-Ulanowicz-Norden-1990-IntJSystSci}
\begin{equation}
\label{eq_redundancy}
\begin{aligned}
r
&= -\sum\limits_{i=1}^{N_\mathcal{V}}\sum\limits_{j=1}^{N_\mathcal{V}} p_{ij}\ln p_{j\mid i}
-\sum\limits_{i=1}^{N_\mathcal{V}}\sum\limits_{j=1}^{N_\mathcal{V}} p_{ij}\ln p_{i\mid j}
\\
&=-\sum\limits_{i=1}^{N_\mathcal{V}}\sum\limits_{j=1}^{N_\mathcal{V}} p_{ij}\ln \frac{p_{ij}}{p_{i}^{\mathrm{out}}}-\sum\limits_{i=1}^{N_\mathcal{V}}\sum\limits_{j=1}^{N_\mathcal{V}} p_{ij}\ln\frac{p_{ij}}{p_{j}^{\mathrm{in}}}
\\
&=\sum_{i=1}^{N_\mathcal{V}}\sum_{j=1}^{N_\mathcal{V}}\frac{f_{ij}}{s}\ln\frac{s_i^{\mathrm{out}}s_j^{\mathrm{in}}}{f_{ij}^2}
\end{aligned},
\end{equation}
where $p_{j\mid i}$ denotes the proportion of the export volume to economy $j$ given that the flow is from economy $i$. Similarly, $p_{i\mid j}$ denotes the proportion of the import volume from economy $i$ given that the flow is to economy $j$. The larger value occurs when trade flows can select more pathways due to the higher uncertainty (i.e., $p_i^{\mathrm{out}}p_j^{\mathrm{in}}\gg p_{ij}^2$ for links with larger volume).

\subsection{Resilience of the international food trade system}

It is well established that improving resilience requires a balance between efficiency (or flexibility) and redundancy (or diversity) \citep{FJ-Zhu-Bao-Qin-Sun-Shia-Chen-2025-AnnOperRes,FJ-Ataburo-Ampong-Essuman-2024-AnnOperRes,FJ-vanStaveren-2023-JEconIssues,FJ-Essuman-Boso-Annan-2020-IntJProdEcon,FJ-Brede-deVries-2009-PhysLettA,FJ-Kamalahmadi-Shekarian-Parast-2022-IntJProdRes}. Based on these two system properties, the ratio $\alpha$, a more comprehensive metric to indicate the order of a system, is proposed for reflecting the trade-off between efficiency and redundancy in the food trade system, which is expressed as
\begin{equation}
    \alpha=\frac{e}{e+r},
    \label{eq_alpha}
\end{equation}
where $0\leq\alpha\leq1$. Inspired by ecological systems that their orders are all close to $1/{\mathrm{e}}$ \citep{FJ-Zorach-Ulanowicz-2003-Complexity}, \cite{FJ-Ulanowicz-2009-EcolModel} defined the fitness $F$ of a system for change to be the product of $\alpha$ and the Boltzmann measure of its disorder such that $F=-c\alpha\ln\alpha$, where $c$ is an appropriate scalar constant. When $\alpha=1/{\mathrm{e}}$, $F'=0$, which means the order of the system is optimized. The underlying assumption is that ecosystems exhibit superior trade-off because they have undergone long-term natural selection \citep{FJ-Ulanowicz-2009-EcolModel,FJ-Liang-Yu-Kharrazi-Fath-Feng-Daigger-Chen-Ma-Zhu-Mi-Yang-2020-NatFood,FJ-Kharrazi-Rovenskaya-Fath-Yarime-Kraines-2013-EcolEcon}. Subsequently, let $c=1$, the resilience $R$ of a food trade network can be defined as
\begin{equation}
    R=-\alpha\ln\alpha,
    \label{eq_resilience_alpha}
\end{equation}
where the maximum value is $1/{\mathrm{e}}\approx 0.3679$. Hence, a resilient food trade system possesses both the capacity to flexibly allocate food (efficient) and to absorb shocks via several pathways (redundant). When $\alpha>1/{\mathrm{e}}$, the system is more efficient and productive but more vulnerable, and vice versa it is more redundant but more inefficient. Notably, resilience vanishes when $\alpha=0$ (overly redundant) or $\alpha=1$ (overly efficient), and more details are provided in \ref{Appendix_SpecialCases}.

\subsection{Relative contributions of changes in efficiency and redundancy}

Given that resilience is constituted by efficiency and redundancy, we can employ the index decomposition analysis to investigate the relative contributions of changes in these two properties to changes in resilience \citep{FJ-Ang-Zhang-2000-Energy,FJ-Liang-Yu-Kharrazi-Fath-Feng-Daigger-Chen-Ma-Zhu-Mi-Yang-2020-NatFood}. We can rewrite Eq.~(\ref{eq_resilience_alpha}) as
\begin{equation}\label{eq_resilience_y}
    R=-\frac{e}{e+r}\ln\left(\frac{e}{e+r}\right).
\end{equation}

Changes in resilience with time $t$ can be defined as
\begin{equation}\label{eq_resilienceTotime}
    \frac{\mathrm{d}R}{\mathrm{d}t}=\frac{1+\ln(e)-\ln(e+r)}{\left(e+r\right)^2}\left(e\frac{\mathrm{d}r}{\mathrm{d}t}-r\frac{\mathrm{d}e}{\mathrm{d}t}\right).
\end{equation}

As we can use $\Delta R$, $\Delta e$, and $\Delta r$ to substitute $\mathrm{d}R$, $\mathrm{d}e$, and $\mathrm{d}r$, respectively, Eq.~(\ref{eq_resilienceTotime}) can be rewritten as
\begin{equation}\label{eq_Delta_R}
\begin{aligned}
    \Delta R &=\frac{1+\ln(e)-\ln(e+r)}{\left(e+r\right)^2}(e\Delta r-r\Delta e)
    \\
    &=\left(\frac{1+\ln(e)-\ln(e+r)}{\left(e+r\right)^2}e\Delta r\right)-\left(\frac{1+\ln(e)-\ln(e+r)}{\left(e+r\right)^2}r\Delta e\right)
    \\
    &\triangleq\Delta_rR+\Delta_eR
\end{aligned},
\end{equation}
where $\Delta_eR$ and $\Delta_rR$ refer to the contributions of changes in efficiency and redundancy to changes in resilience, respectively. 

As only discrete data are available in empirical studies, we approximate $e$ and $r$ by the arithmetic mean of their respective values for base year and observation year \citep{FJ-Ang-Zhang-2000-Energy}. Subsequently, we can define a ratio as follows \citep{FJ-Liang-Yu-Kharrazi-Fath-Feng-Daigger-Chen-Ma-Zhu-Mi-Yang-2020-NatFood}:
\begin{equation}
    \mathrm{ratio}\equiv\frac{\Delta_{r}R}{\Delta_{e}R}=-\frac{e\Delta r}{r\Delta e}=-\frac{\frac{1}{2}(e_t+e_0)\Delta r}{\frac{1}{2}(r_t+r_0)\Delta e}=-\frac{(e_t+e_0)(r_t-r_0)}{(r_t+r_0)(e_t-e_0)}.
\end{equation}
As we can calculate efficiency $e_t$ and redundancy $r_t$ at any time $t$, $\Delta_eR$ and $\Delta_rR$ can be calculated by
\begin{equation}
    \Delta_eR=(R_t-R_0)\frac{\Delta_eR}{\Delta_rR+\Delta_eR}=(R_t-R_0)\frac{1}{1+\mathrm{ratio}}
    \label{eq_Delta_eR}
\end{equation}
and
\begin{equation}
    \Delta_rR=(R_t-R_0)\frac{\Delta_rR}{\Delta_rR+\Delta_eR}=(R_t-R_0)\frac{\mathrm{ratio}}{1+\mathrm{ratio}}.
    \label{eq_Delta_rR}
\end{equation}

\subsection{Relative contributions of changes in P, PMI, and PMR}

Furthermore, we decompose resilience into three internal components and investigate the contributions of changes in these components to changes in resilience \citep{FJ-Liang-Yu-Kharrazi-Fath-Feng-Daigger-Chen-Ma-Zhu-Mi-Yang-2020-NatFood}. The first internal component is the concentration degree of trade flows, defined as a P matrix containing element $p_{ij}$ that is expressed as Eq.~(\ref{eq_p_ij}), where a higher $p_{ij}$ indicates a greater trade flow relative intensity.

The second internal component is the inter-dependency among economies, defined as a point-wise mutual information (PMI) matrix containing element $\mathrm{pmi}_{ij}$ that denotes the degree of inter-dependency between economies $i$ and $j$ \citep{FJ-Kharrazi-Fath-2016-EP}. A higher value indicates a higher probability of a flow between given economies. The element $\mathrm{pmi}_{ij}$ of PMI matrix is expressed as
\begin{equation}
    \mathrm{pmi}_{ij}=\ln\frac{p_{ij}}{p_{i}^{\mathrm{out}}p_{j}^{\mathrm{in}}}
    =\ln\frac{f_{ij}s}{s_i^{\mathrm{out}}s_j^{\mathrm{in}}}.
    \label{eq_pmi}
\end{equation}

According to Eq.~(\ref{eq_efficiency}), we start by calculating contributions of changes in $p_{ij}$ and $\mathrm{pmi}_{ij}$ to changes in efficiency by the index decomposition method \citep{FJ-Ang-Zhang-2000-Energy}:
\begin{equation}
\begin{aligned}
    \Delta e_{ij}
    &=\Delta p_{ij}\times\mathrm{pmi}_{ij}+p_{ij}\times\Delta\mathrm{pmi}_{ij}
    \\
    &=\frac{1}{2}\left(\Delta p_{ij}\times\mathrm{pmi}_{ij}^0+\Delta p_{ij}\times\mathrm{pmi}_{ij}^t\right)+\frac{1}{2}\left(p_{ij}^0\times\Delta\mathrm{pmi}_{ij}+p_{ij}^t\times\Delta\mathrm{pmi}_{ij}\right)
    \\
    &=\Delta p_{ij}\times\frac{1}{2}\left(\mathrm{pmi}_{ij}^0+\mathrm{pmi}_{ij}^t\right)+\Delta\mathrm{pmi}_{ij}\times\frac{1}{2}\left(p_{ij}^0+p_{ij}^t\right)
\end{aligned},
\end{equation}
where $\Delta e_{ij}$ denotes changes in efficiency caused by changes in the trade link from economy $i$ to $j$. Hence, the contributions of changes in P and PMI to changes in efficiency can be written as
\begin{equation}
    \Delta e=\Delta_{p}e+\Delta_\mathrm{pmi}e,
\end{equation}
where
\begin{equation}
    \Delta_{p}e=\sum\limits_{i=1}^{N_\mathcal{V}}\sum\limits_{j=1}^{N_\mathcal{V}}\Delta_{p}e_{ij}=\sum\limits_{i=1}^{N_\mathcal{V}}\sum\limits_{j=1}^{N_\mathcal{V}}\Delta p_{ij}\times\frac{1}{2}\left(\mathrm{pmi}_{ij}^0+\mathrm{pmi}_{ij}^t\right)
\end{equation}
and
\begin{equation}
    \Delta_\mathrm{pmi}e=\sum\limits_{i=1}^{N_\mathcal{V}}\sum\limits_{j=1}^{N_\mathcal{V}}\Delta_\mathrm{pmi}e_{ij}=\sum\limits_{i=1}^{N_\mathcal{V}}\sum\limits_{j=1}^{N_\mathcal{V}}\Delta \mathrm{pmi}_{ij}\times\frac{1}{2}\left(p_{ij}^0+p_{ij}^t\right)
\end{equation}
denote the contributions of changes in the concentration degree of trade flows and those in the inter-dependency among economies to changes in network efficiency, respectively.

The third internal component is the inter-independency among economies, defined as a point-wise mutual redundancy (PMR) matrix containing element $\mathrm{pmr}_{ij}$ that denotes the degree of inter-independency between economies $i$ and $j$ \citep{FJ-Liang-Yu-Kharrazi-Fath-Feng-Daigger-Chen-Ma-Zhu-Mi-Yang-2020-NatFood}. A higher value indicates a higher diversity for the origin and destination of flows between given economies. Similarly, the element $\mathrm{pmr}_{ij}$ of PMR matrix is expressed as
\begin{equation}
    \mathrm{pmr}_{ij}=-\ln p_{j\mid i}-\ln p_{i\mid j}=\ln\frac{s_i^{\mathrm{out}}s_j^{\mathrm{in}}}{f_{ij}^2}.
\end{equation}

Hence, the contributions of changes in P and PMR to changes in redundancy can be written as:
\begin{equation}
    \Delta r=\Delta_{p}r+\Delta_\mathrm{pmr}r,
\end{equation}
where
\begin{equation}
    \Delta_{p}r=\sum\limits_{i=1}^{N_\mathcal{V}}\sum\limits_{j=1}^{N_\mathcal{V}}\Delta_{p}r_{ij}=\sum\limits_{i=1}^{N_\mathcal{V}}\sum\limits_{j=1}^{N_\mathcal{V}}\Delta p_{ij}\times\frac{1}{2}\left(\mathrm{pmr}_{ij}^0+\mathrm{pmr}_{ij}^t\right)
\end{equation}
and
\begin{equation}
    \Delta_\mathrm{pmr}r=\sum\limits_{i=1}^{N_\mathcal{V}}\sum\limits_{j=1}^{N_\mathcal{V}}\Delta_\mathrm{pmr}r_{ij}=\sum\limits_{i=1}^{N_\mathcal{V}}\sum\limits_{j=1}^{N_\mathcal{V}}\Delta \mathrm{pmr}_{ij}\times\frac{1}{2}\left(p_{ij}^0+p_{ij}^t\right)
\end{equation}
denote the contributions of changes in the concentration degree of trade flows and those in the inter-independency among economies to changes in network redundancy, respectively.

Subsequently, we can rewrite Eqs.~(\ref{eq_Delta_eR}) and (\ref{eq_Delta_rR}) as
\begin{equation}
    \Delta_eR=\mathrm{peR}\times\Delta e=\mathrm{peR}\times(\Delta_{p}e+\Delta_\mathrm{pmi}e)
\end{equation}
and
\begin{equation}
    \Delta_rR=\mathrm{prR}\times\Delta r=\mathrm{prR}\times(\Delta_{p}r+\Delta_\mathrm{pmr}r),
\end{equation}
where peR and prR denote changes in resilience caused by unitary changes in efficiency and redundancy, expressed respectively as
\begin{equation}
    \mathrm{peR}=\frac{\Delta_eR}{\Delta e}=\frac{R_t-R_0}{e_t-e_0}\times\frac{1}{1+\mathrm{ratio}}
\end{equation}
and
\begin{equation}
    \mathrm{prR}=\frac{\Delta_rR}{\Delta r}=\frac{R_t-R_0}{r_t-r_0}\times\frac{\mathrm{ratio}}{1+\mathrm{ratio}}.
\end{equation}

Ultimately, we decompose changes in resilience into three components:
\begin{equation}
\begin{aligned}
    \Delta R
    &=\Delta_eR+\Delta_rR
    \\
    &=\mathrm{peR}\times\Delta e+\mathrm{prR}\times\Delta r
    \\
    &=(\mathrm{peR}\times\Delta_{p}e+\mathrm{prR}\times\Delta_{p}r)+(\mathrm{peR}\times\Delta_\mathrm{pmi}e)+(\mathrm{prR}\times\Delta_\mathrm{pmr}r)
    \\
    &\triangleq\Delta_{p}R+\Delta_\mathrm{pmi}R+\Delta_\mathrm{pmr}R
\end{aligned},
\end{equation}
where $\Delta_{p}R$, $\Delta_\mathrm{pmi}R$, and $\Delta_\mathrm{pmr}R$ denote the contributions of changes in P, PMI, and PMR to changes in resilience, respectively.

\subsection{Relative contributions of changes in economies and trade relationships}

The perturbation methods, including node removal, link removal, and the global weight perturbation, are commonly employed to investigate changes in system resilience \citep{FJ-Liu-Xu-Gao-Wang-Li-Gao-2024-NatCommun}. Since the global weight perturbation does not affect results in Eqs.~(\ref{eq_efficiency}) and (\ref{eq_redundancy}), according to the leave-one-out approach \citep{FJ-Hue-Lucotte-Tokpavi-2019-JEconDynControl, FJ-Zedda-Cannas-2020-JBankFinanc}, we employ node removal and link removal to quantify the contributions of changes in economies and relationships to changes in resilience, thereby identifying critical economies and trade relationships from the resilience perspective. 

We denote the resilience of the networked food trade system $G$ with $N_\mathcal{V}$ economies and $N_\mathcal{E}$ trade relationships as $R^G$. For a given economy $i$, we consider a system $G_{(-i)}$ with $N_\mathcal{V}-1$ nodes that includes all economies except economy $i$ and denote the resilience of this system as $R^{G_{(-i)}}$. Hence, the contributions of changes in economy $i$ can be expressed as
\begin{equation}\label{eq_deltaRes_i}
    \Delta R_i=\frac{\lvert R^G-R^{G_{(-i)}}\rvert}{R^G},
\end{equation}
which also denotes the significance of economy $i$ in the international food trade system from a network resilience perspective. A higher $\Delta R_i$ implies that economy $i$ plays a more important role in system resilience.

Similarly, for a given trade relationship $\varepsilon_{ij}$, we consider a system $G_{(-\varepsilon_{ij})}$ with $N_\mathcal{E}-1$ links that includes all relationships except the relationship $\varepsilon_{ij}$ and denote the resilience of this system as $R^{G_{(-\varepsilon_{ij})}}$. Hence, the contributions of changes in relationship $\varepsilon_{ij}$ can be expressed as
\begin{equation}
    \Delta R_{\varepsilon_{ij}}=\frac{\lvert R^G-R^{G_{(-\varepsilon_{ij})}}\rvert}{R^G},
    \label{eq_deltaRes_e_ij}
\end{equation}
which also denotes the significance of relationship $\varepsilon_{ij}$ in the international food trade system from a network resilience perspective. A higher $\Delta R_{\varepsilon_{ij}}$ implies that relationship $\varepsilon_{ij}$ plays a more important role in system resilience.

\subsection{Resilience of core and peripheral food trade networks}

The international food trade network exhibits a pronounced core-periphery structure \citep{FJ-Dong-Yin-Lane-Yan-Shi-Liu-Bell-2018-PhysA}, which prompts us to investigate the differences between resilience of core and peripheral food trade networks. It should be emphasized that our study explicitly excludes cumulative economy removal scenarios, as termination of trade relationships remains operationally feasible \citep{FJ-Guo-Chen-Fan-Zeng-Pu-Qing-2025-ExpertSystAppl}, which fundamentally differs from complete embargoes on economy's food trade. We sort all trade relationships either in ascending or descending order of $\Delta R_{\varepsilon_{ij}}$.
In our analysis, a proportion $p$ of the relationships are first removed from the ordered set and $G_{(-p)}$ is defined as the food trade network that includes the remaining proportion $1-p$ of relationships in order of $\Delta R_{\varepsilon_{ij}}$. For ascending order of $\Delta R_{\varepsilon_{ij}}$, the network $G_{(-p)}$ can be regarded as a core food trade network when the value of $p$ is large (that is, the network is constituted by relationships with greater impact on resilience). Also, for descending order of $\Delta R_{\varepsilon_{ij}}$, the network $G_{(-p)}$ can be regarded as a peripheral food trade network when the value of $p$ is large (that is, the network is constituted by relationships with less impact on resilience). Subsequently, we can calculate the resilience of core or peripheral food trade networks $R^{G_{(-p)}}$ with respect to the changes in $p$, which may help clarify the role of the above networks in sustaining network resilience and guide structure optimization for the international food trade.

\section{Dynamic evolution of the international food trade network and its resilience}
\label{Sec_Evolution}

Given that maize, rice, soybean, and wheat are foods that account for most of the calories consumed by the world's population \citep{FJ-DOdorico-Carr-Laio-Ridolfi-Vandoni-2014-EarthsFuture, FJ-Zhang-Zhou-2023-ChaosSolitonFract}, we select the international bilateral trade quantities of maize (CPC code\footnote{Central Product Classification (CPC) code, a product classification for goods and services promulgated by the United Nations Statistics Division.}: 0112), rice (CPC code: F0030), soybean (CPC code: 0141), and wheat (CPC code: 0111) from 1986 to 2022, provided by the ``detailed trade matrix'' dataset\footnote{FAOSTAT, \url{www.fao.org/faostat/en/\#data/TM}.} of the FAO, to construct networks, where economies are considered as nodes and trade relationships between economies serve as directed and weighted links with the trade quantities being the weights of links. We note that the abbreviations for economies used in the figures and tables are taken from the International Standard Organization\footnote{ISO, \url{www.iso.org/iso-3166-country-codes.html}.}. 

\begin{figure}[htb!]
\centering 
\includegraphics[width=1\textwidth]{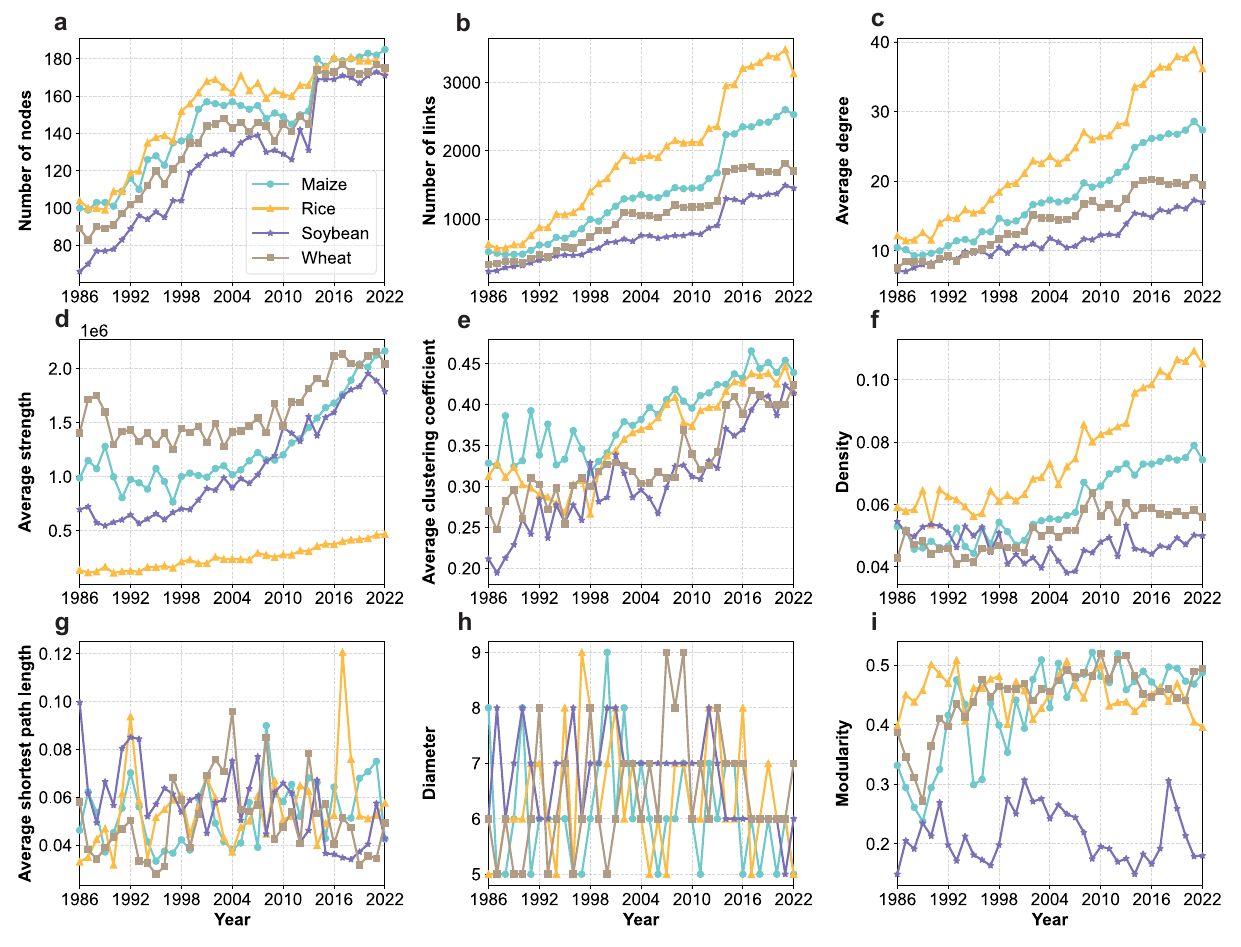} 
\caption{The evolution of topological indicators of international food trade networks. {\bf{a,b}}, The numbers of nodes $N_{\mathcal{V}}$ and links $N_{\mathcal{E}}$ are two alternative measures of network size. {\bf{c,d}}, Average degree $\langle k\rangle$ and average strength $\langle s\rangle$ represent the average number of trade partners and trade volume of economies, respectively. {\bf{e}}, Average clustering coefficient $\langle C\rangle$ represents the average actual connections among the neighbors of individual nodes. {\bf{f}}, Density $\rho$ reflects the level of interconnectedness among economies. {\bf{g}}, Average shortest path length $L$ represents the level of separation among economies. {\bf{h}}, Diameter $D$ represents the maximum of shortest path lengths between any two economies. {\bf{i}}, Modularity $Q$ represents the denseness of flows within a community compared to between communities. }\label{Plot_Characteristic_1986_2022_withLabels}
\end{figure}

Generally, there is a trend of increasing network size (Fig.~\ref{Plot_Characteristic_1986_2022_withLabels}a,b). The size of rice trade network is the largest, followed by that of the maize, wheat, and soybean trade networks. However, the average strength for rice is always lower than that of other foods due to its low trade volume (Fig.~\ref{Plot_Characteristic_1986_2022_withLabels}d). In 2022, for example, the total quantities traded globally for maize, soybean, and wheat are 180 million tons, 168 million tons and 217 million tons, respectively, compared to only 55 million tons for rice. Different from other foods with industrial production attributes, rice is predominantly edible and has a homogenous consumption structure. Also, high self-sufficiency on the rice-producing side and entrenched regional consumption habits\footnote{For example, China imports only 2\% of its rice consumption in broken rice \citep{FJ-Harold-Ashok-Valerien-Takashi-David-2024-GlobFoodSecur}.} result in low trade volume for rice \citep{ FJ-Lakkakula-Dixon-Thomsen-Wailes-Danforth-2015-AgricEcon}. Other reasons are analyzed later. It is worth noting that the number of economies participating in global trade has increased significantly in 2014 with some African and Oceanian economies joining trade networks, driven by a surge in demand from other developing economies. For example, Papua New Guinea and Mozambique, relying on resources of undeveloped arable land, have for the first time exported foods such as maize to large importers\footnote{OECD-FAO Agricultural Outlook 2013-2022, \url{https://openknowledge.fao.org/handle/20.500.14283/ar251e}.}. 

\begin{figure}[htb!]
\centering 
\includegraphics[width=1\textwidth]{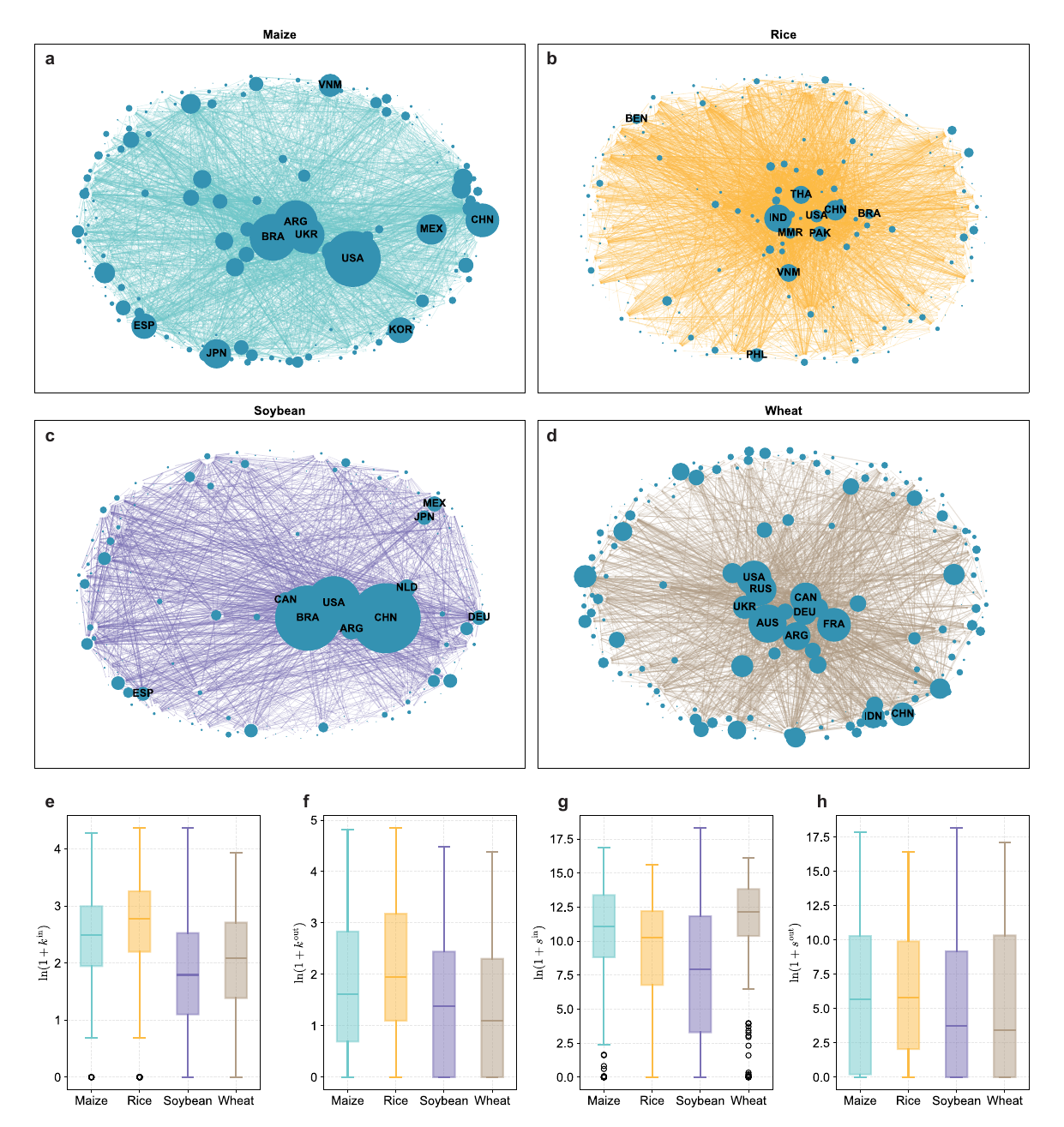} 
\caption{The international trade networks of maize, rice, soybean, and wheat in 2022. {\bf{a-d}}, The international trade network of maize $(N_\mathcal{V}=185, N_\mathcal{E}=2531)$, rice $({N_\mathcal{V}}=173, N_\mathcal{E}=3130)$, soybean $({N_\mathcal{V}}=171, N_\mathcal{E}=1453)$, and wheat $({N_\mathcal{V}}=175, N_\mathcal{E}=1703)$ in 2022. {\bf{e-h}}, The logarithmic-transformed boxplots of four indicators in 2022, including in-degree ($k^{\mathrm{in}}$), out-degree ($k^{\mathrm{out}}$), in-strength ($s^{\mathrm{in}}$), and out-strength ($s^{\mathrm{out}}$).}\label{Plot_Combined_Network_Distribution_2022}
\end{figure}

Meanwhile, we find that the clustering coefficients of the international food trade networks are increasing  (Fig.~\ref{Plot_Characteristic_1986_2022_withLabels}e) and the average shortest pathway length of the international access to foods fluctuates markedly (Fig.~\ref{Plot_Characteristic_1986_2022_withLabels}g), especially during food crises (e.g., 07/08 and 11/12) and around major events (e.g., the dissolution of the Soviet Union in 1991 and the COVID-19 pandemic since 2020). In addition, since relatively large modularity values between 0.3 and 0.7 are considered an indicator of a well-structured network partition \citep{FJ-Newman-2004-PhysRevE}, we find that flows within communities are denser compared to flows between communities for maize, rice, and wheat (Fig.~\ref{Plot_Characteristic_1986_2022_withLabels}i).

Specifically, we analyze the international trade networks of four foods in 2022, as shown in Fig.~\ref{Plot_Combined_Network_Distribution_2022}. The node size refers to its total trade volume (including imports and exports) and we label the top 10 economies by the size (Fig.~\ref{Plot_Combined_Network_Distribution_2022}a-d). Evidently, the United States and China are the most important trading economies, with the former being a major exporter and the latter a major importer. With regard to exports, the United States has vast arable land and suitable climatic condition for the large-scale cultivation of a wide range of foods. Not only in production but also in variety, the adequate level of supply has also contributed to it being a major exporter. With regard to imports, although China has a large area of arable land, it has a large population and a small per capita area of arable land, which causes China having to rely on imports to meet domestic demands. 

Compared to maize and rice, the soybean and wheat trade networks exhibit sparser connectivity and lower density (Fig.~\ref{Plot_Characteristic_1986_2022_withLabels}f), manifested by comparable numbers of participating economies but fewer trade relationships (Fig.~\ref{Plot_Characteristic_1986_2022_withLabels}a,b). For both of them, the first quartile is zero in the logarithmic-transformed boxplots of $k^{\mathrm{out}}$ and $s^{\mathrm{out}}$ (Fig.~\ref{Plot_Combined_Network_Distribution_2022}f,h), indicating their more concentrated exporting characteristics. The difference, however, is that trade volume of wheat varies less among importers and is more evenly distributed (Fig.~\ref{Plot_Combined_Network_Distribution_2022}g). Moreover, the soybean trade network exhibits substantial heterogeneity in the size of participating economies, indicating highly monopolistic characteristics by a few economies such as the United States, Brazil, and China (Fig.~\ref{Plot_Combined_Network_Distribution_2022}c). 

\begin{figure}[htb!]
\centering 
\includegraphics[width=1\textwidth]{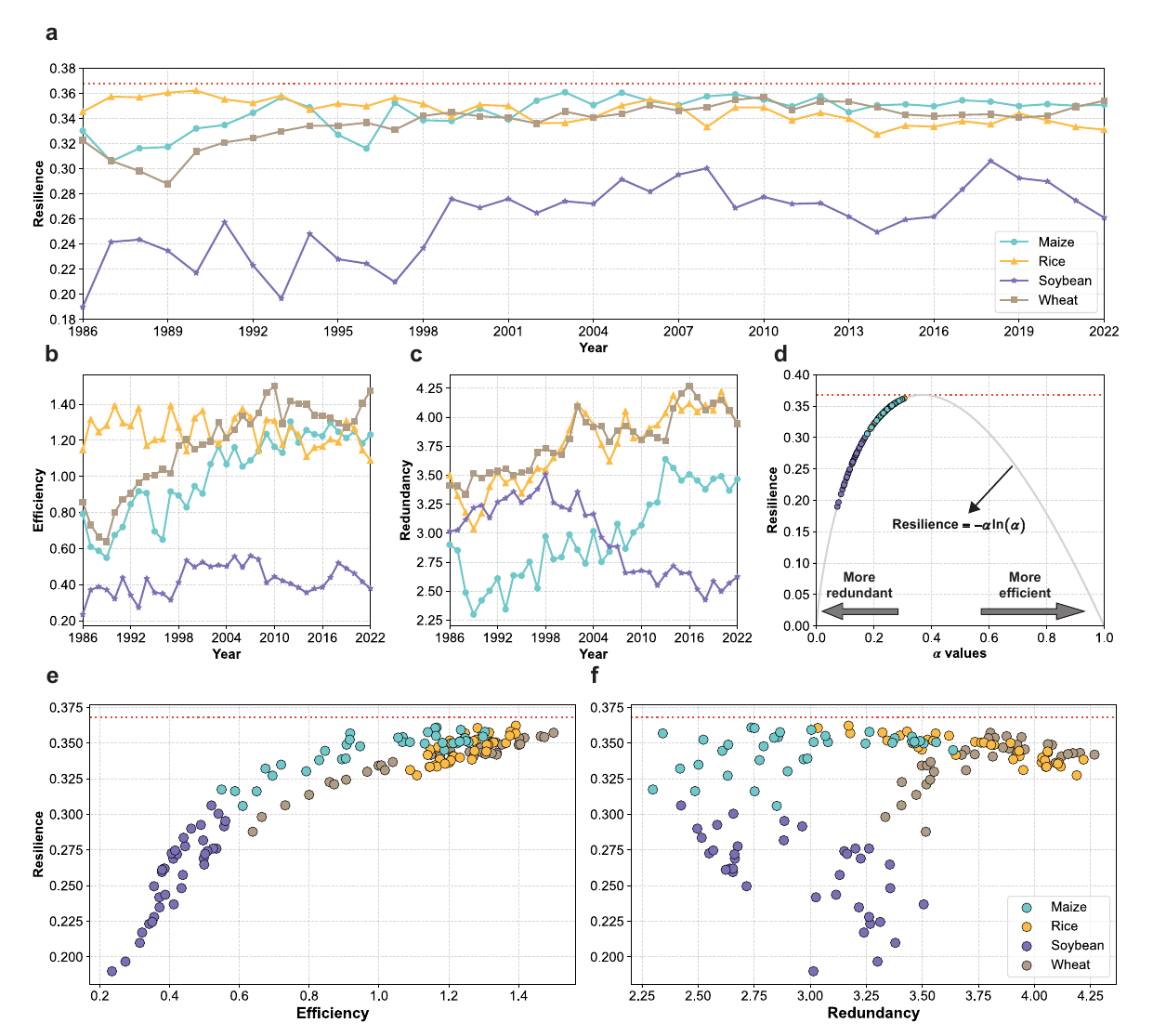} 
\caption{The evolution of several properties of international food trade networks. The resilience {\bf{(a)}}, efficiency {\bf{(b)}}, redundancy {\bf{(c)}}, and $\alpha$ values {\bf{(d)}} of international food trade networks from 1986 to 2022. In this figure, the red dotted line indicates $1/{\mathrm{e}}\approx0.3679$. {\bf{e,f}}, The correlations between resilience and two properties: efficiency and redundancy.}
\label{Plot_Combined_EffRedAlpRes_withLabels}
\end{figure}

Furthermore, we calculate efficiency, redundancy, $\alpha$ values, and resilience of the international food trade networks from 1986 to 2022 (Fig.~\ref{Plot_Combined_EffRedAlpRes_withLabels}a-d). Overall, trade networks of maize, rice, and wheat are more resilient. Separately, maize and wheat exhibit parallel enhancements in both efficiency and redundancy, resulting in the better trade-off within the $\alpha$ framework. Rice exhibits enhancement in redundancy but fluctuation in efficiency, resulting in a slowly declining resilience in recent years. Conversely, the soybean trade network is not only more vulnerable, its resilience fluctuates more significantly. As is known, approximately 87\% of soybean export quantities originated from Brazil and the United States in 2022. Due to the excessively high concentration of exporters, most economies, except for the main exporters, remain relatively isolated, with limited alternative trade pathways. Consequently, the soybean trade network lacks the capacity for development and both efficiency and redundancy, and exhibits an insufficient trade-off within the $\alpha$ framework, ultimately resulting in lower resilience. Notably, $\alpha$ values of all networks are less than $1/{\mathrm{e}}$, resulting in the highly positive correlation between efficiency and resilience (Fig.~\ref{Plot_Combined_EffRedAlpRes_withLabels}e) and the negative correlation between redundancy and resilience (Fig.~\ref{Plot_Combined_EffRedAlpRes_withLabels}f). It indicates that the current trade networks possess sufficient redundancy but require enhanced efficiency, especially for soybean.

\section{Structural drivers of resilience of the international food trade network}
\label{Sec_Correlation}

\subsection{Correlations between topological indicators and resilience}

\begin{figure}[htb!]
\centering 
\includegraphics[width=1\textwidth]{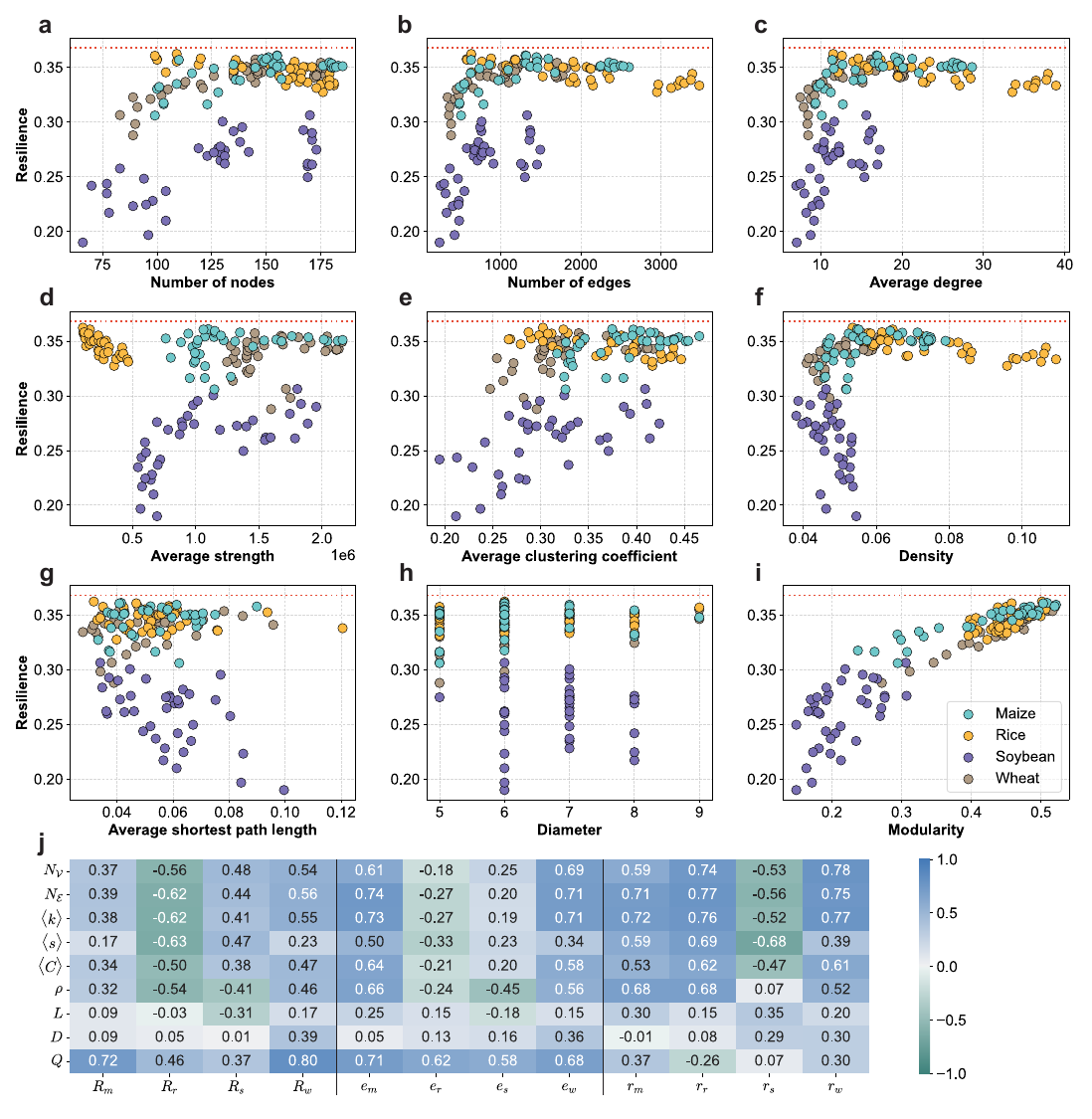} 
\caption{{\bf{a-i}}, Scattered distribution of resilience and topological indicators. In this figure, the red dotted line indicates $1/{\mathrm{e}}$. {\bf{j}}, Kendall's correlation coefficients between resilience and topological indicators, between efficiency and topological indicators, and between redundancy and topological indicators. In this figure, $R_m$, $R_r$, $R_s$, and $R_w$ denote resilience of trade networks of maize, rice, soybean, and wheat, respectively. $e_m$, $e_r$, $e_s$, and $e_w$ denote efficiency of trade networks of maize, rice, soybean, and wheat, respectively. $r_m$, $r_r$, $r_s$, and $r_w$ denote redundancy of trade networks of maize, rice, soybean, and wheat, respectively.}
\label{Plot_Topological_Resilience_withLabels}
\end{figure}

We investigate the correlations between topological indicators of the international food trade networks and their resilience and also calculate the correlation coefficients between topological indicators and both efficiency and redundancy, as shown in Fig.~\ref{Plot_Topological_Resilience_withLabels}. 

Across all staple foods, we find that modularity exhibits a stronger positive correlation with both efficiency and resilience due to the current networks being more redundant (more details are provided in \ref{Appendix_Modularity}). This findings can also explain why more resilient states of international trade are associated with stronger community structures \citep{FJ-Garlaschelli-Loffredo-2005-PhysA, FJ-Li-Wang-Kharrazi-Fath-Liu-Liu-Xiao-Lai-2024-FoodSecur}. For maize and wheat, for example, the correlation coefficients between resilience and modularity are 0.72 and 0.80, while those between efficiency and modularity are 0.71 and 0.68, respectively (Fig.~\ref{Plot_Topological_Resilience_withLabels}j). A higher modularity reflects denser intra-community flows relative to inter-community flows, corresponding to processes of regional integration that lead to stronger inter-dependence among economies within a community and, in turn, enhanced overall network efficiency \citep{FJ-deRaymond-Alpha-BenAri-Daviron-Nesme-Tetart-2021-GlobFoodSecur}. Within our framework, the resilience of the current networks is positively correlated with their efficiency, so higher modularity implies higher resilience.

The correlation between resilience and other topological indicators varies across different foods. For example, the clustering characteristics are positively correlated with resilience for maize, soybean, and wheat. According to Fig.~\ref{Plot_Topological_Resilience_withLabels}j, in trade networks such as those for maize and wheat, evident clustering characteristics enhance trade efficiency (e.g., through regional alliances and international cooperation organizations, such as NAFTA\footnote{NAFTA, The North American Free Trade Area.} and MERCOSUR\footnote{MERCOSUR, The Southern Common Market.}, and visualizations are provided in \ref{Appendix_Community}). Together with modularity, these features jointly contribute to sustaining network resilience. However, resilience of the rice trade network appears to decline with increases in both clustering coefficients and the network density, potentially due to its inherent network characteristics and the nature of rice as a commodity. On one hand, the rice network is heavily concentrated around a few major exporters (such as India, Thailand, and Vietnam) and importers (such as China and the Philippines), forming a single-core structure of Asia. Despite denser connectivity, this over-centralized architecture has exacerbated network vulnerability. Meanwhile, while geographic proximity suggests potential trade alliances, the lack of strategic complementarity, exemplified by competitive underpricing between Vietnam and Thailand, undermines cluster resilience by fostering internal fragmentation. Additionally, as an inelastic staple commodity with low substitution elasticity, the rice trade network faces catastrophic systemic collapse once major Asian producers implement export restrictions to ensure domestic food security (e.g., India's 2022 export ban for rice).

\subsection{Relative contributions of changes in efficiency and redundancy to changes in resilience}

\begin{figure}[htb!]
\centering 
\includegraphics[width=1\textwidth]{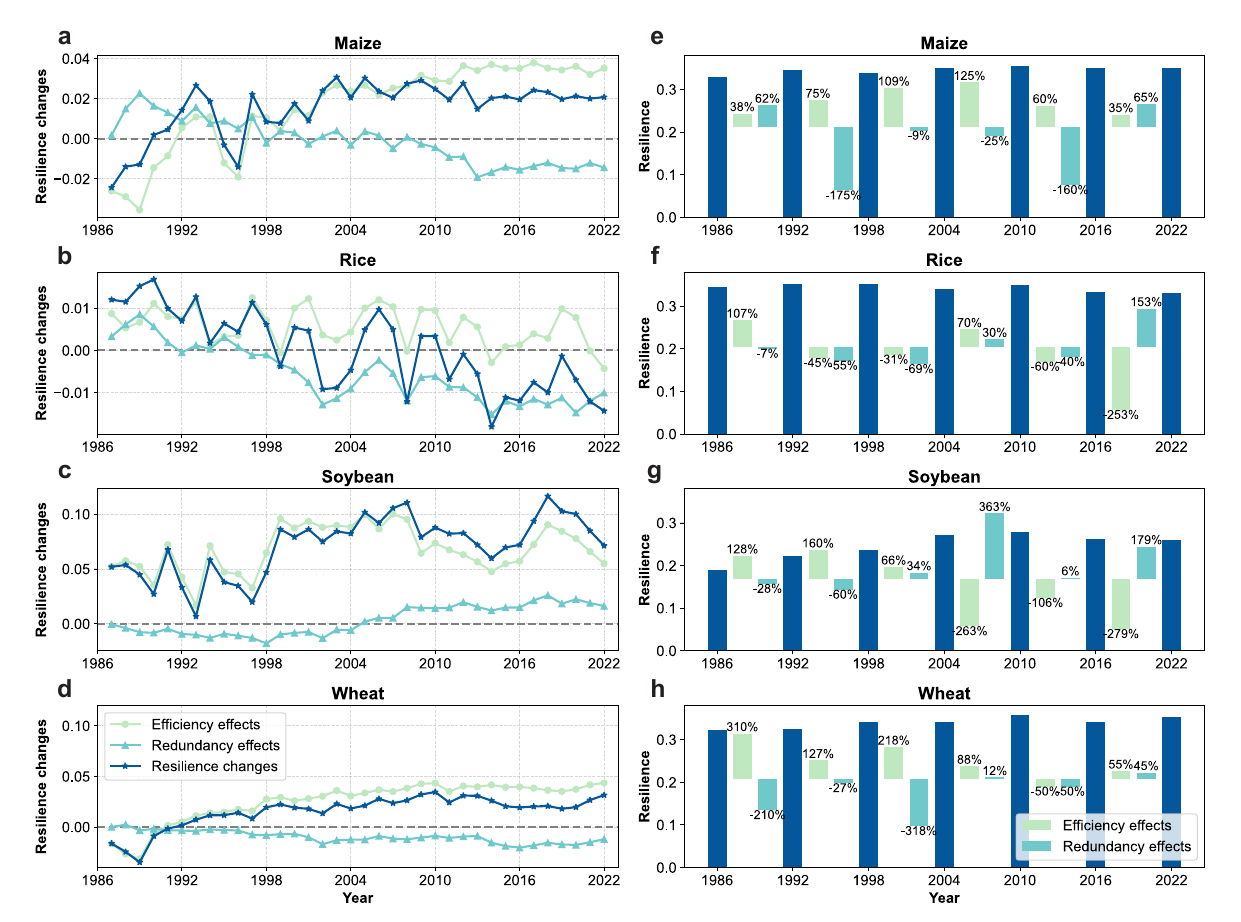} \caption{Relative contributions of changes in efficiency and redundancy to changes in resilience of international food trade networks for 1986-2022 ({\bf{a-d}}) and specific time periods ({\bf{e-h}}).} \label{Plot_Bar_EffRed_Resilience_withLabels}
\end{figure}

We calculate the relative contributions of changes in efficiency and redundancy to changes in network resilience for 1986-2022 (Fig.~\ref{Plot_Bar_EffRed_Resilience_withLabels}a-d). As previously mentioned, the current trade networks are relatively redundant due to $\alpha<1/{\mathrm{e}}$. Hence, changes in resilience has been dominated by changes in efficiency. For example, although the redundancy effects in the soybean trade network are lower than those in the base year prior to 2004, the network resilience is also higher during this period because of continued positive changes in efficiency (Fig.~\ref{Plot_Bar_EffRed_Resilience_withLabels}c). The exception is rice. Before 1998, changes in resilience of the rice trade network has been dominated by changes in efficiency. However, the effect has been diverse after 2010 with the dominance of redundancy (Fig.~\ref{Plot_Bar_EffRed_Resilience_withLabels}b), and the reason (increasing redundancy and oscillating efficiency) can also be found in Fig.~\ref{Plot_Combined_EffRedAlpRes_withLabels}b.

We also calculate the relative contributions of changes in efficiency and redundancy to changes in network resilience for specific time periods (Fig.~\ref{Plot_Bar_EffRed_Resilience_withLabels}e-h). Different from what is found above, we find that changes in resilience are not always dominated by one property or another. For example, the 0.03 increase (from 0.19 to 0.22) in resilience of the soybean trade network during 1986 to 1992 is due to the 128\% increase in efficiency, but the 0.006 increase (from 0.272 to 0.278) in resilience of that during 2004 to 2010 is due to the 363\% increase in redundancy (Fig.~\ref{Plot_Bar_EffRed_Resilience_withLabels}g). Moreover, the 0.02 increase (from 0.32 to 0.34) in resilience of the wheat trade network during 1992 to 1998 is due to the 127\% increase in efficiency, but the 0.02 decrease (from 0.342 to 0.34) in resilience of the wheat trade network during 1998 to 2004 is due to the 318\% decrease in redundancy (Fig.~\ref{Plot_Bar_EffRed_Resilience_withLabels}h). The findings suggest that the dominance of changes in efficiency to changes in resilience is diminishing while $\alpha$ values are closer to $1/{\mathrm{e}}$.

\subsection{Relative contributions of changes in P, PMI, and PMR to changes in resilience}

\begin{figure}[htb!]
\centering 
\includegraphics[width=1\textwidth]{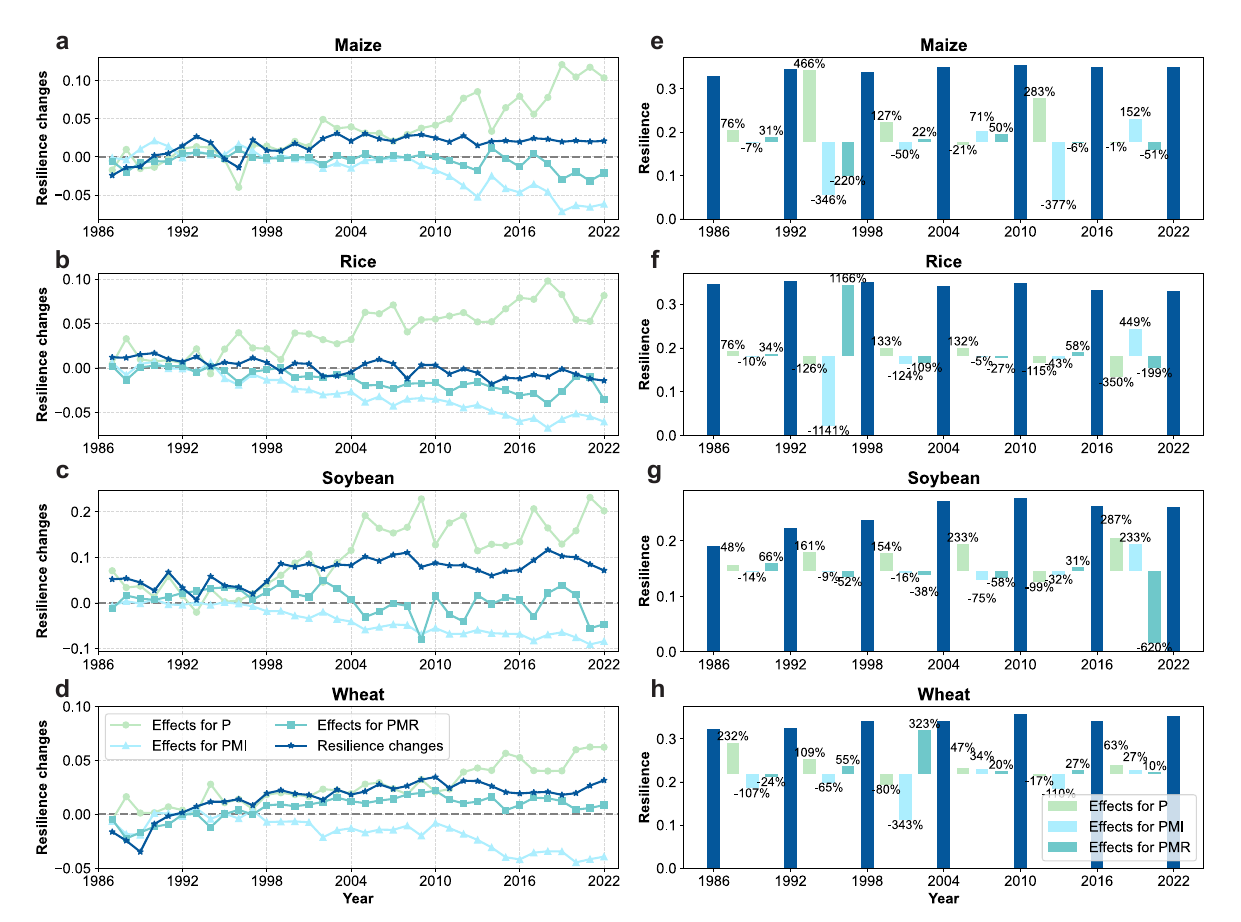} \caption{Relative contributions of changes in the concentration degree of trade flows (P), economies inter-dependency (PMI), and economies inter-independency (PMR) to changes in resilience of international food trade networks for 1986-2022 ({\bf{a-d}}) and specific time periods ({\bf{e-h}}).} \label{Plot_Bar_P_PMI_PMR_Resilience_withLabels}
\end{figure}

Furthermore, we calculate the relative contributions of changes in the concentration degree of trade flows, economies inter-dependency, and economies inter-independency to changes in network resilience for 1986-2022 (Fig.~\ref{Plot_Bar_P_PMI_PMR_Resilience_withLabels}a-d). Overall, changes in the concentration degree of trade flows contribute more to changes in resilience for maize, soybean, and wheat throughout the entire period of 1986-2022, indicating the greater influence of trade flows concentration for food demand on resilience \citep{FJ-Liang-Yu-Kharrazi-Fath-Feng-Daigger-Chen-Ma-Zhu-Mi-Yang-2020-NatFood}. However, these effects are diminishing with trade globalization, resulting in negative impact on changes in resilience. Indeed, these findings reflect the underlying shift in food trade discourse from flow monopolies to strategic interactions within relational networks. For example, resilience for rice has been dominated by changes in the concentration degree of trade flows until 2010. Since then, effects of relationships among economies has been highlighted, resulting in decreasing resilience. Although the effect for the concentration degree of trade flows of the rice trade network is higher than 5\% after 2010, changes in resilience exhibit phase inversion due to synergistic interaction for interrelationships (Fig.~\ref{Plot_Bar_P_PMI_PMR_Resilience_withLabels}b).

We also calculate the relative contributions of changes in the concentration degree of trade flows, economies inter-dependency, and economies inter-independency to changes in network resilience for specific time periods (Fig.~\ref{Plot_Bar_P_PMI_PMR_Resilience_withLabels}e-h). We find that changes in resilience are dominated by the concentration degree of trade flows and economies inter-dependency, corroborating prior observations that efficiency-driven mechanisms exert predominant influence. Exceptionally, the 0.001 decrease (from 0.262 to 0.261) in resilience of the soybean trade network during 2016 to 2022 is due to the 620\% decrease in economies inter-independency (Fig.~\ref{Plot_Bar_P_PMI_PMR_Resilience_withLabels}g).

\section{Identifying critical economies and trade relationships from the resilience perspective}
\label{Sec_Identify}

\subsection{Relative contributions of changes in economies and trade relationships to changes in resilience}

\begin{figure}[htb!]
\centering 
\includegraphics[width=1\textwidth]{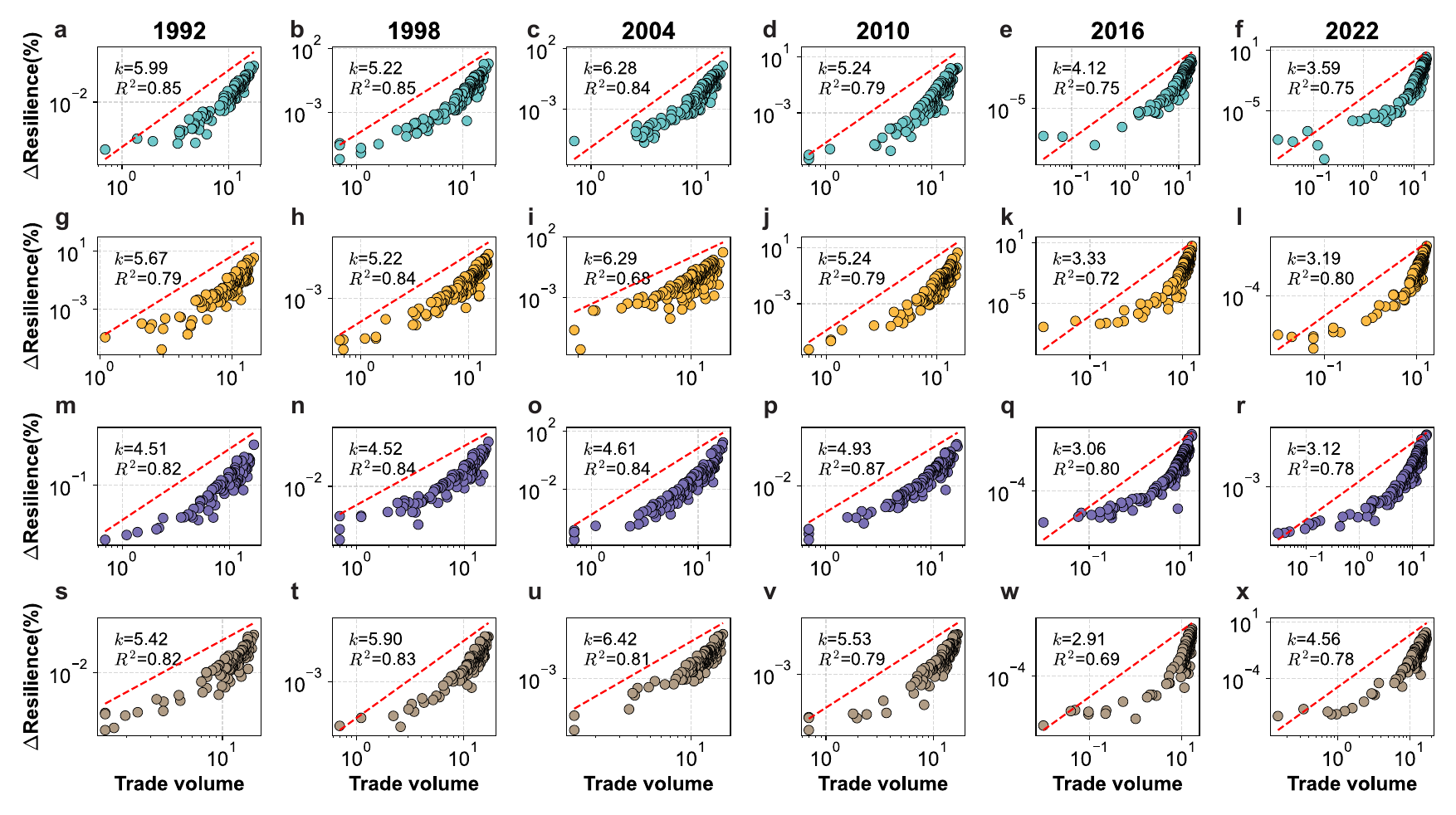} \caption{The correlations between trade volume of economy $i$ and changes in resilience of the international food trade network as the economy $i$ is removed during 1992 to 2022. In this figure, the red dashed line denotes the fit to scatters, where $k$ is slope and $R^2$ is goodness of fit. {\bf{a-f}}, The maize trade networks. {\bf{g-l}}, The rice trade networks. {\bf{m-r}}, The soybean trade networks. {\bf{s-x}}, The wheat trade networks.}
\label{Plot_Scatter_DropNodes_DeltaResilience_withLabels}
\end{figure}

\begin{figure}[htb!]
\centering 
\includegraphics[width=1\textwidth]{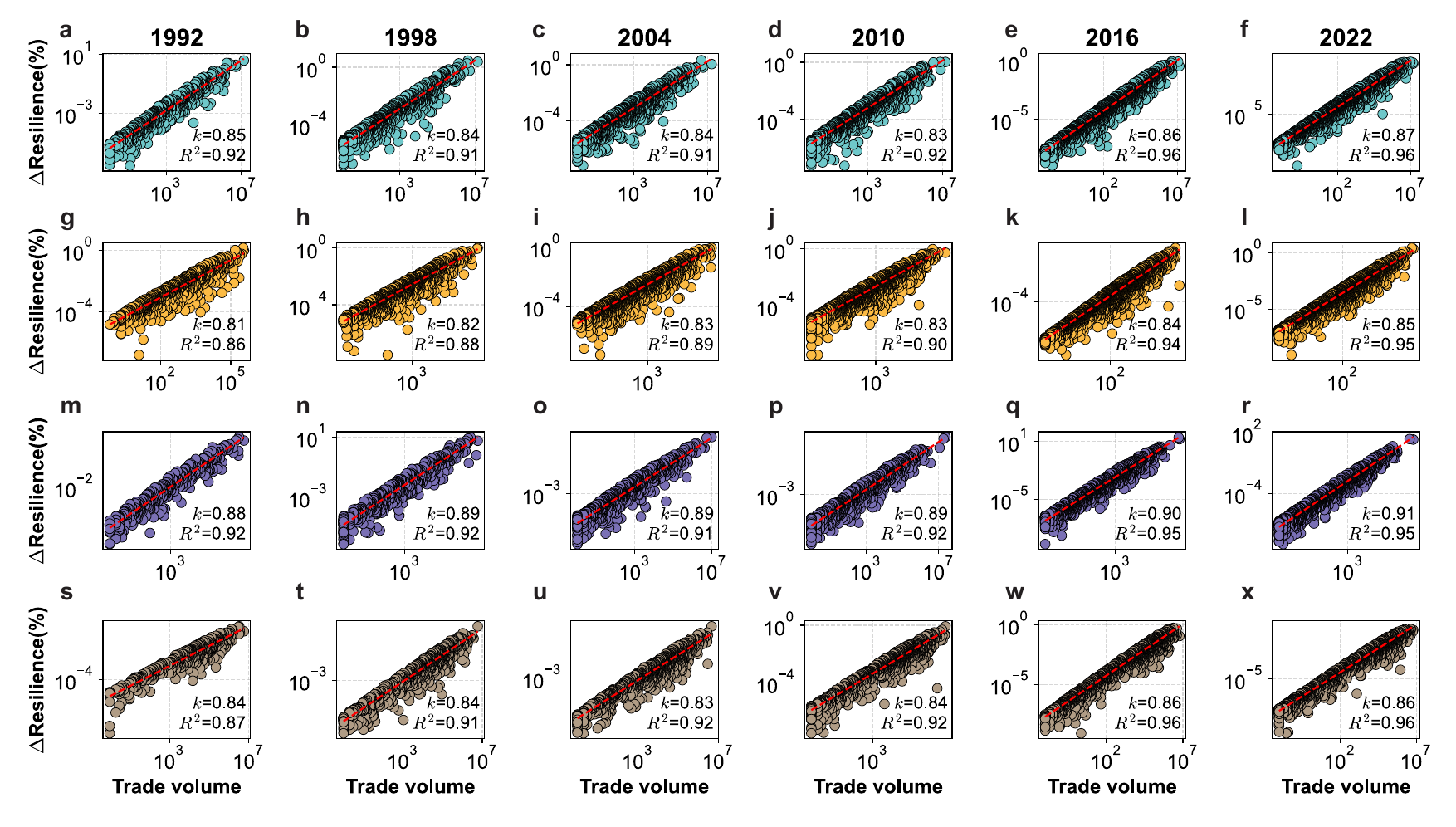} \caption{The correlations between trade volume from economy $i$ to $j$ and changes in resilience of the international food trade network as the relationship $\varepsilon_{ij}$ is removed during 1992 to 2022. In this figure, the red dashed line denotes the fit to scatters, where $k$ is slope and $R^2$ is goodness of fit. {\bf{a-f}}, The maize trade networks. {\bf{g-l}}, The rice trade networks. {\bf{m-r}}, The soybean trade networks. {\bf{s-x}}, The wheat trade networks.}
\label{Plot_Scatter_DropEdges_DeltaResilience_withLabels}
\end{figure}

To quantify the relative contributions of changes in economies and trade relationships to changes in resilience, following the leave-one-out approach, we remove some given economy or trade relationship independently and regard changes in the network resilience ($\Delta R$) as its relative contribution. Figs.~\ref{Plot_Scatter_DropNodes_DeltaResilience_withLabels} and \ref{Plot_Scatter_DropEdges_DeltaResilience_withLabels} respectively depict the log-log correlations between $\Delta R$ and trade volume of both economy and relationship during 1992 to 2022.

Overall, trade volume is significantly positively correlated with contributions to resilience, indicating that economies or trade relationships with larger volumes are generally more critical for sustaining network resilience. For economies, however, we find that the slopes in recent years are significantly smaller than those in earlier years. For rice, for example, the slopes exceeded 5.2 during 1992-2010 but decreased to 3.3 after 2016. By contrast, the slopes for trade relationships have remained relatively stable and even increased slightly. These results suggest that the influence of economies' trade volumes on their resilience contributions is declining, while the role of interactions is stable and becoming more important compared to flow monopolies, which are also consistent with our decomposition analysis of the relative contributions of changes in P, PMI, and PMR (Fig.~\ref{Plot_Bar_P_PMI_PMR_Resilience_withLabels}).

Meanwhile, the scale-free characteristics of the international food trade system are reaffirmed from the perspective of changes in resilience caused by economies and trade relationships. That is, a small number of economies and trade relationships contribute most of the changes in resilience, and their influences on resilience demonstrates a power-law distribution. For example, in 1998, only four economies and a single trade relationship contribute more than 2\% to resilience in the international wheat trade network, while all remaining economies and relationships each contributes less than 1\%\footnote{The results are available on request due to space limitations.}. Over time, the disparity in participants' contributions to network resilience has narrowed, particularly in the maize and wheat trade networks, indicating that mature networks possess more balanced structures. For example, while four economies and a single trade relationship contribute more than 2.5\% to resilience of the international maize trade network in 1992 and 1998, only a single economy contributes more than 2\% to that in 2022, as well as all trade relationships contributing less than 1\%. This finding suggests that more economies and trade relationships are integrating in the system and lessening the significance of the dominators and thus making the system more balanced and resilient, aligning with previous findings \citep{FJ-Sartori-Schiavo-2015-FoodPol}.

To further investigate the heterogeneity among leading contributors, we respectively present the 2022 top 10 economies and trade relationships in Tables~\ref{Table_Economies} and \ref{Table_Relationships}, based on their relative contributions to resilience. Overall, the contributions of leading economies and trade relationships to resilience are highest in the soybean trade network, followed by those in rice, maize, and wheat, which aligns with the resilience characteristics discussed above. It shows that the top 10 economies' $\Delta R$ values range from 1.64\% to 17.06\% for soybean, compared to 0.82\%-5.79\% for rice, 0.62\%-2.30\% for maize, and 0.34\%-1.14\% for wheat. Also, the top 10 trade relationships’ $\Delta R$ values range from 1.75\% to 23.59\% for soybean, 0.70\%-2.55\% for rice, 0.49\%-0.74\% for maize, and 0.23\%-0.31\% for wheat. These findings suggest that the soybean trade network exhibits substantially greater vulnerability to disruptions compared to its maize and wheat counterparts.

\afterpage{
\begin{sidewaystable}
\renewcommand\arraystretch{0.9}
\renewcommand{\tabcolsep}{1mm}
\caption{The top 10 economies in 2022 by relative contributions to the international food trade network resilience and their centrality measures values.}
\label{Table_Economies}
{\centering
\footnotesize
    \begin{tabular}{lrrrrrrrrrrrrrrrrrrrrrrrrrrrrr}
\toprule
    ~ & \multicolumn{7}{c}{Maize} & \multicolumn{7}{c}{Rice} & \multicolumn{7}{c}{Soybean} & \multicolumn{7}{c}{Wheat}  \\
    \cmidrule(lr){2-8}\cmidrule(lr){9-15}\cmidrule(lr){16-22}\cmidrule(lr){23-29}
    $\mathrm{R}_r$ & Econ & $\Delta R$ & $\mathrm{R}_v$ & $\mathrm{R}_{dc}$ & $\mathrm{R}_{bc}$ & $\mathrm{R}_{pc}$ & $\mathrm{R}_{ec}$ & Econ & $\Delta R$ & $\mathrm{R}_v$ & $\mathrm{R}_{dc}$ & $\mathrm{R}_{bc}$ & $\mathrm{R}_{pc}$ & $\mathrm{R}_{ec}$ & Econ & $\Delta R$ & $\mathrm{R}_v$ & $\mathrm{R}_{dc}$ & $\mathrm{R}_{bc}$ & $\mathrm{R}_{pc}$ & $\mathrm{R}_{ec}$ & Econ & $\Delta R$ & $\mathrm{R}_v$ & $\mathrm{R}_{dc}$ & $\mathrm{R}_{bc}$ & $\mathrm{R}_{pc}$ & $\mathrm{R}_{ec}$ \\
    \midrule
    1 & BRA(e)  & 2.30\%  & 2  & 15  & 43  & 98 & 142 & IND(e)  & 5.79\%  & 1  & 4  &  10  & 35 & 83 & BRA(e)  & 17.06\%  & 2 & 12 & 44 & 98 & 147 & CAN(e)  & 1.14\%  & 5 & 5 & 6 & 39 & 23 \\ 
    2 & UKR(e)  & 1.45\%  & 4 & 18  & 38  & 101 & 35 & CHN(i)  & 2.81\%  & 2  &  13  &  27  & 47 & 101 & CHN(i)  & 16.89\%  & 1 & 5 & 8 & 12 & 44 & RUS(e)  & 0.98\%  & 4  & 14 & 15 & 37 & 29 \\ 
    3 & USA(e)  & 1.44\%  & 1 & 2 & 3 & 18 & 26 & THA(e)  & 2.69\%  & 3  &  7  &  16  & 110 & 68 & USA(e)  & 13.44\%  & 3 & 2 & 3 & 13 & 29 & FRA(e)  & 0.71\%  & 3 & 2 & 2 & 21 & 4 \\ 
    4 & KOR(i)  & 1.26\%  & 8 & 46 & 57 & 42 & 47 & PAK(e)  & 2.39\%  & 5  &  14  &  21  & 125 & 94 & ARG(e)  & 7.87\%  & 4 & 22 & 23 & 47 & 85 & KAZ(e)  & 0.64\%  & 23  & 31 & 25 & 62 & 54 \\ 
    5 & JPN(i)  & 0.98\%  & 7 & 42 & 32 & 29 & 33 & BRA(i)  & 2.31\%  & 9  & 31  &  53  & 56 & 106 & PRY(e)  & 7.13\%  & 20 & 57 & 83 & 110 & 157 & IDN(i)  & 0.59\%  & 10 & 67 & 76 & 52 & 82  \\ 
    6 & PRY(e)  & 0.75\%  & 22 & 76 & 129 & 172 & 171 & PHL(i)  & 2.02\%  & 6 & 51 & 69 & 79 & 95 & UKR(e)  & 2.78\%  & 23 & 14 & 19 & 65 & 27 & UKR(e)  & 0.58\%  & 9 & 10 & 26 & 150 & 40  \\ 
    7 & IND(i)  & 0.68\%  & 30 & 13 & 10 & 82 & 82 & VNM(e)  & 1.74\%  & 4  & 15 & 125 & 143 & 143 & DEU(i)  & 2.40\%  & 8 & 7 & 9 & 2 & 3 & TUR(i)  & 0.49\%  & 11 & 13 & 12 & 10 & 21 \\ 
    8 & ESP(i)  & 0.68\%  & 9 & 11 & 13 & 11 & 10 & PRY(i)  & 1.46\%  & 28 & 50 & 63 & 63 & 99 & IND(i)  & 2.08\%  & 34 & 8 & 6 & 15 & 68 & ALG(i)  & 0.43\%  & 16 & 56 & 97 & 17 & 37 \\ 
    9 & COL(i)  & 0.62\%  & 14 & 71 & 40 & 41 & 144 & MOZ(i)  & 0.84\%  & 13 & 80 & 36 & 23 & 57 & BOL(e)  & 1.79\%  & 32 & 72 & 102 & 140 & 161 & BRA(i)  & 0.39\%  & 12 & 36 & 24 & 34 & 76 \\ 
    10 & ARG(e)  & 0.62\%  & 3 & 5 & 12 & 128 & 84 & USA(i)  & 0.82\%  & 7 & 2 & 2 & 8 & 15 & ESP(i)  & 1.64\%  & 10  & 13 & 13 & 3 & 7 & ITA(i)  & 0.34\%  & 14 & 6 & 13 & 7 & 7 \\  
    \bottomrule
\end{tabular}}
\begin{flushleft}
    \footnotesize
Notes: Econ denotes the economy's name. $\Delta R$ denotes the economy's relative contribution to network resilience. $\mathrm{R}_v$, $\mathrm{R}_{dc}$, $\mathrm{R}_{bc}$, $\mathrm{R}_{pc}$, $\mathrm{R}_{ec}$, and $\mathrm{R}_r$ denote the economy's ranks by trade volume, degree centrality \citep{FJ-Freeman-1978-SocNetwork}, betweenness centrality \citep{FJ-Freeman-1977-Sociometry}, PageRank centrality \citep{FJ-Brin-Page-1998-ComputNetwISDNSyst}, eigenvector centrality \citep{FJ-Bonacich-1972-JMathSoc}, and $\Delta R$, respectively. The export or import property of economy is given in parentheses.
\end{flushleft}

\vspace{0.5\baselineskip}

\renewcommand{\tabcolsep}{1.3mm}
\caption{The top 10 relationships in 2022 by relative contributions to the international food trade network resilience.}
\label{Table_Relationships}
{\centering
    \begin{tabular}{lrrrrrrrrrrrrrrrr}
    \toprule
        ~ & \multicolumn{4}{c}{Maize} & \multicolumn{4}{c}{Rice} & \multicolumn{4}{c}{Soybean} & \multicolumn{4}{c}{Wheat}  \\
       \cmidrule(lr){2-5}\cmidrule(lr){6-9}\cmidrule(lr){10-13}\cmidrule(lr){14-17}
        $\mathrm{R}_r$ & Exporter & Importer & $\Delta R$ & $\mathrm{R}_v$ & Exporter & Importer & $\Delta R$ & $\mathrm{R}_v$ & Exporter & Importer & $\Delta R$ & $\mathrm{R}_v$ & Exporter & Importer & $\Delta R$ & $\mathrm{R}_v$
        \\ \midrule
        1 & USA & JPN  & 0.74\%  & 3 & IND & CHN  & 2.55\%  & 2 & USA & CHN  & 23.59\%  & 2 & AUS & IDN  & 0.31\%  & 4 \\ 
        2 & BRA & JPN  & 0.69\%  & 10 & VNM & PHL  & 2.15\%  & 1 & BRA & CHN  & 20.95\%  & 1 & ARG & BRA  & 0.31\%  & 3 \\ 
        3 & UKR & CHN  & 0.65\%  & 6 & VNM & CHN  & 1.25\%  & 5 & PRY & ARG  & 4.64\%  & 16 & FRA & CHN  & 0.30\%  & 20 \\ 
        4 & USA & CHN  & 0.63\%  & 2 & THA & CHN  & 1.22\%  & 10 & NLD & DEU  & 2.50\%  & 23 & CAN & CHN  & 0.27\%  & 19 \\ 
        5 & PRY & BRA  & 0.62\%  & 19 & PRY & BRA  & 1.21\%  & 17 & BRA & NLD  & 2.49\%  & 12 & RUS & TUR  & 0.26\%  & 1 \\ 
        6 & BRA & ESP  & 0.56\%  & 9 & PAK & CHN  & 0.97\%  & 3 & URY & CHN  & 2.15\%  & 13 & CAN & IDN  & 0.25\%  & 26 \\ 
        7 & USA & MEX  & 0.51\%  & 1 & IND & CIV  & 0.84\%  & 14 & USA & NLD  & 1.89\%  & 15 & FRA & EGY  & 0.24\%  & 31 \\ 
        8 & ARG & KOR  & 0.50\%  & 5 & IND & SEN  & 0.76\%  & 6 & BRA & ESP  & 1.87\%  & 9 & FRA & ALG  & 0.24\%  & 13 \\ 
        9 & BRA & KOR  & 0.50\%  & 20 & IND & BEN  & 0.75\%  & 4 & CAN & CHN  & 1.80\%  & 28 & CZE & DEU  & 0.24\%  & 22 \\ 
        10 & USA & KOR  & 0.49\%  & 28 & IND & MOZ  & 0.70\%  & 20 & BRA & TUR  & 1.75\%  & 10 & AUS & PHL  & 0.23\%  & 8 \\   \bottomrule
    \end{tabular}
    }
\begin{flushleft}
    \footnotesize
Notes: $\Delta R$ denotes the relationship's relative contribution to network resilience, expressed as Eq.~(\ref{eq_deltaRes_e_ij}). $\mathrm{R}_v$ and $\mathrm{R}_r$ denote the relationship's ranks by trade volume and $\Delta R$, respectively.
\end{flushleft}
\end{sidewaystable}
}

In terms of economies, we find that with the exception of China, which is the main importer, the top 3 contributors are all exporters, including Brazil, Ukraine, the United States, India, Thailand, Canada, Russia, and France. This underscores the greater influence of an economy’s export role on network resilience \citep{FJ-Grassia-Mangioni-Schiavo-Traverso-2022-SciRep}. Generally, export restrictions raise the risk of supply chain disruptions and trigger cascading effects that undermine the resilience \citep{FJ-Burkholz-Schweitzer-2019-EnvironResLett}. For example, maize exports from Ukraine have been disrupted by the Russia-Ukraine conflict, which not only destabilizes maize imports in other economies but also further shocks the supply of pork and poultry meat due to the lack of fodder maize \citep{FJ-Laber-Klimek-Bruckner-Yang-Thurner-2023-NatFood}. Meanwhile, export restrictions tend to disrupt the global food price and the resulting price volatility induces the trade resilience. For example, after Ukraine's food exports are disrupted, 18 economies impose food export bans, leading to a 23\% increase in the global food price index. In contrast, import restrictions primarily impact unilateral market access, whereas exporters can mitigate risks through trade partner diversification and inventory expansion, thereby exerting only a limited impact on overall trade resilience. 

Although the results exhibits the positive correlations between trade volume and resilience contributions, according to Tables~\ref{Table_Economies} and \ref{Table_Relationships}, the two should not be directly equated. In the case of economies, for a comprehensive comparison, we also present their rankings based on degree centrality, betweenness centrality, PageRank centrality, and eigenvector centrality, four of the most commonly used centrality measures in the network science literature \citep{FJ-Li-Liao-Yen-2013-ResPolicy,FJ-Lu-Zhou-Zhang-Stanley-2016-NatCommun,FJ-Lambiotte-Rosvall-Scholtes-2019-NatPhys,FJ-Liu-Xiong-Perera-Guo-2024-IntJProdRes,FJ-GutierrezMoya-AdensoDiaz-Lozano-2021-FoodSecur,FJ-Zhang-Zhou-2023-ChaosSolitonFract} in Table~\ref{Table_Economies}. Among them, degree centrality captures the importance of an economy in terms of the number of its trade partners, whereas betweenness centrality reflects its role as a bridge in the network, measured by the number of shortest paths that pass through it. PageRank centrality and eigenvector centrality emphasize the importance of an economy depends on both the number and the importance of its neighboring economies. 

We find that the rankings based on trade volume and four centrality measures capture more accurately the strategic importance of economies such as Spain in the maize trade network, the United States in the rice trade network, Germany in the soybean trade network, and Italy in the wheat trade network. These economies are developed and interact with many partners, thereby occupying both core and bridging positions in the networks. However, such measures may fail to capture the role of some developing producers and exporters, including India, Thailand, and the Philippines in the rice trade network, Brazil in the soybean trade network, and Ukraine in the wheat trade network. These economies are evidently important due to their export capacity and grain yields, yet their trade partners are typically consumers rather than re-exporters. As a result, PageRank and eigenvector centrality often classify them as peripheral nodes, thereby overlooking their strategic importance. Consequently, we propose that $\Delta R$ can complement some conventional centrality measures, as illustrated by the following two cases.

Paraguay ranks sixth (0.75\%), eighth (1.46\%), and fifth (7.13\%) in contributions to resilience for maize, rice, and soybean, respectively, despite falling outside the top 20 in trade volume and the top 50 in four centrality measures for all three commodities. Three potential explanations may account for this phenomenon. First, serving as a strategic redundancy reserve in South America's multimodal corridors, Paraguay leverages its geopolitical advantage as a landlocked state to activate the Paraguay-Paran\'{a} Waterway intermodal system during congestion mitigation protocols at primary Brazilian ports (e.g., Santos). This enables maize rerouting through alternative terminals like Paranagu\'{a}, sustaining 15\% of displaced shipments while maintaining supply chain continuity (see results of $\Delta R=0.62\%$ for Paraguay-Brazil maize trade relationship in Table~\ref{Table_Relationships}). Second, to address the demands of Brazil's premium culinary market, Paraguay operationalizes a ``low-volume and high-premium'' trade paradigm by importing Thai Jasmine rice for value-added processing before re-exporting to Brazil. This strategic intervention mitigates production shortfalls in domestic indica rice cultivation. It is further substantiated by the elevated $\Delta R$ values in Paraguay-Brazil and Paraguay-Argentina, as quantified in Table~\ref{Table_Relationships}. Third, Paraguay stands as the sole nation maintaining 100\% Non-GMO\footnote{Non-GMO, Products that are produced without genetic engineering.} soybean commercial cultivation globally, positioning itself to fulfill the European Union's stringent Non-GMO demand (evidenced by a 58\% export surge in 2022), thereby securing an irreplaceable supply chain role. 

Similarly, Kazakhstan ranks fourth (0.64\%) in contributions to resilience for wheat, despite falling outside the top 20 in trade volume and all centrality measures. Its critical role stems from functioning as a crisis replacement corridor, offering alternative routes, such as the China-Europe Railway and the Trans-Caspian Corridor, when the Black Sea passage or conventional land routes are disrupted (e.g., the Russia-Ukraine conflict). Moreover, Kazakhstan supplies 65\% of China's organic wheat imports for animal feed, fulfilling the demand for ``Non-GMO and high-gluten'' products, thereby securing an indispensable position in the international wheat trade system. 

Cases such as Paraguay and Kazakhstan demonstrate that strategic trade positioning may surpass the indiscriminate pursuit of trade volume in enhancing contributions to the international food trade resilience, which may be an indication for other economies.

\subsection{Comparing resilience of core and peripheral food trade networks}

\begin{figure}[htb!]
\centering 
\includegraphics[width=1\textwidth]{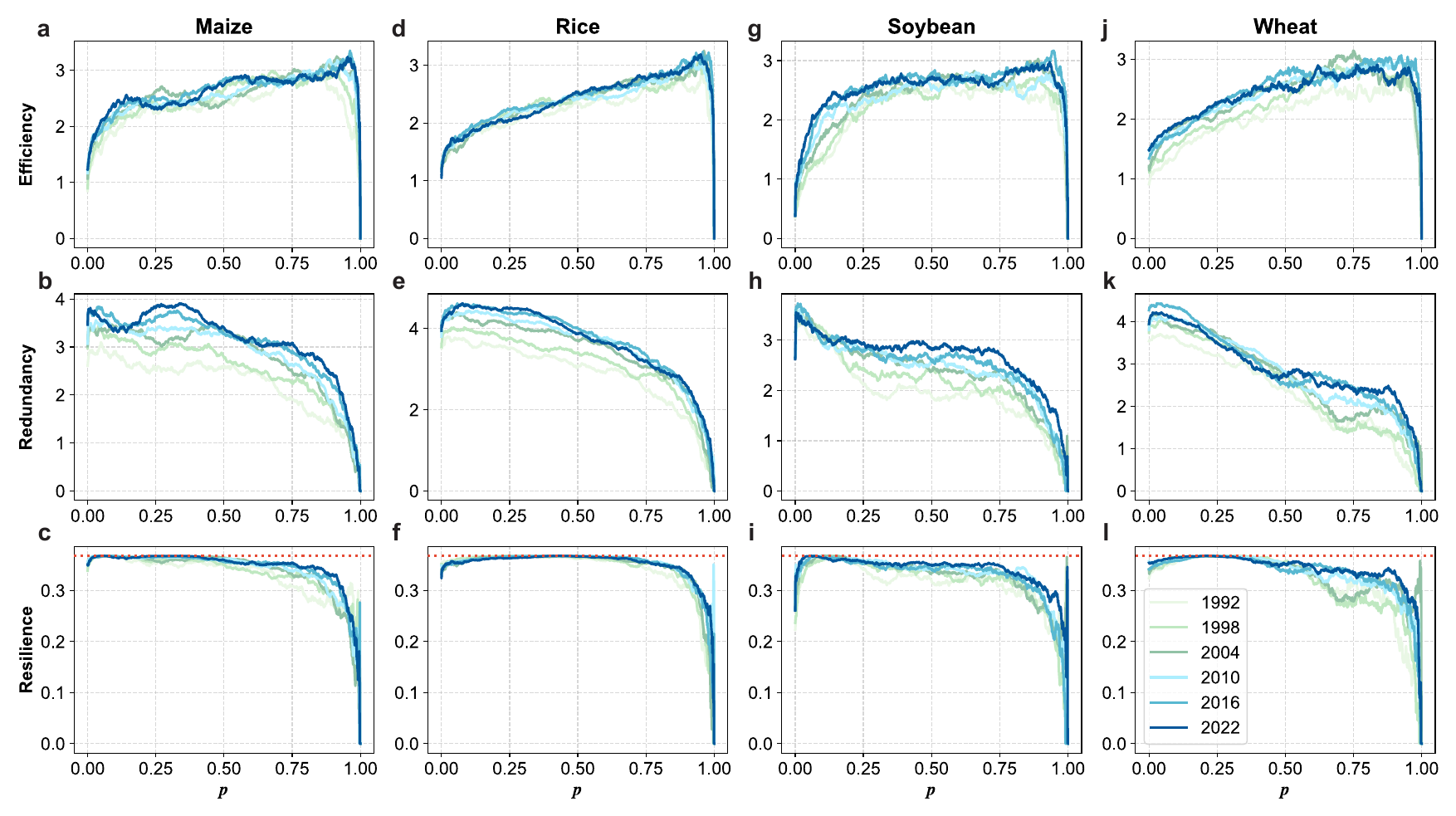} 
\caption{The evolution of efficiency, redundancy, and resilience of the international food trade sub-networks as trade relationships are removed cumulatively in descending order of $\Delta R$. The red dotted line indicates $1/{\mathrm{e}}$. {\bf{a-c}}, The maize trade networks. {\bf{d-f}}, The rice trade networks. {\bf{g-i}}, The soybean trade networks. {\bf{j-l}}, The wheat trade networks.}\label{Plot_Resilience_DropEdges_Descend_withLabels}
\end{figure}

\begin{figure}[htb!]
\centering
\includegraphics[width=1\textwidth]{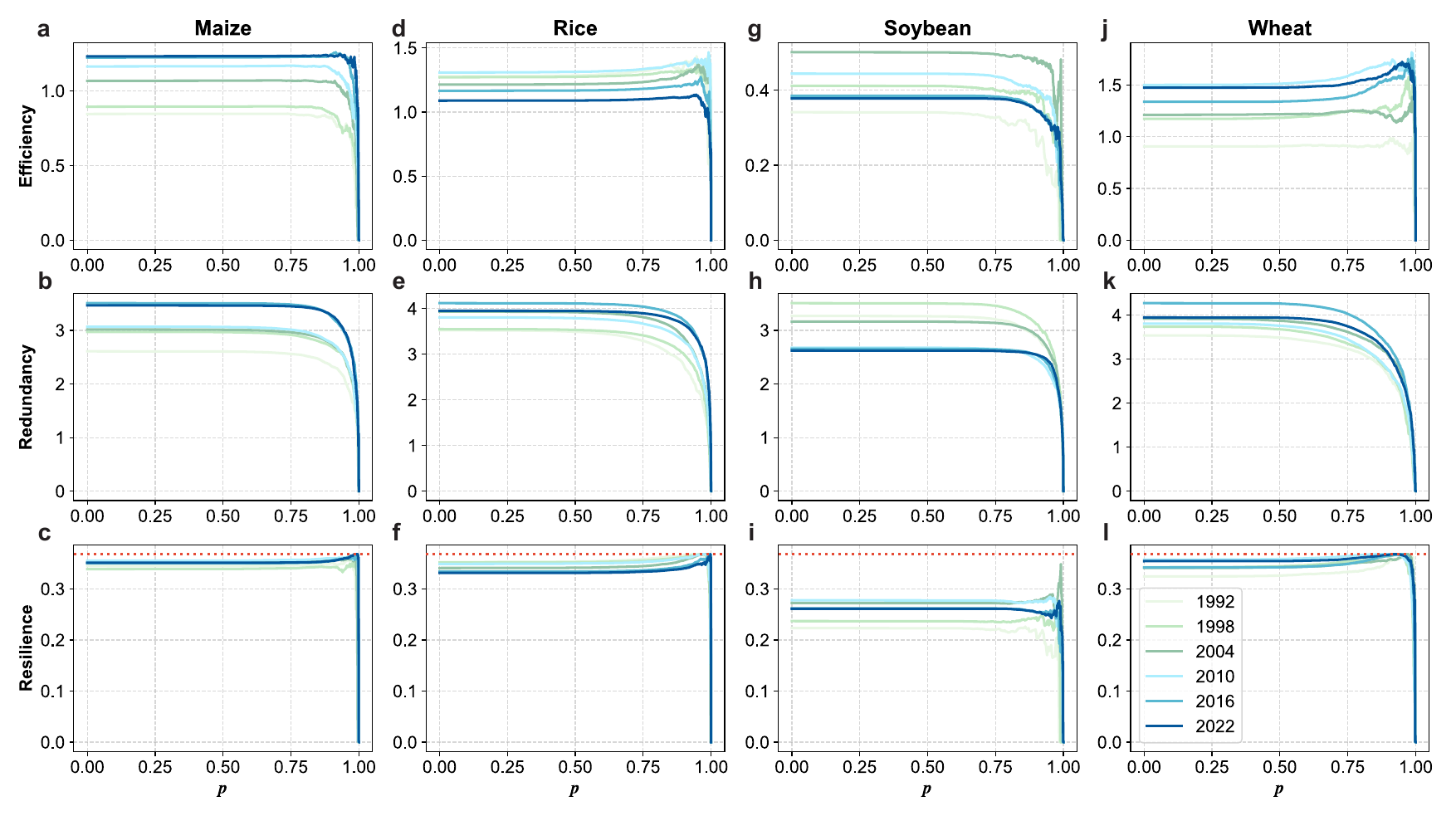} 
\caption{The evolution of efficiency, redundancy, and resilience of the international food trade sub-networks as trade relationships are removed cumulatively in ascending order of $\Delta R$. The red dotted line indicates $1/{\mathrm{e}}$. {\bf{a-c}}, The maize trade networks. {\bf{d-f}}, The rice trade networks. {\bf{g-i}}, The soybean trade networks. {\bf{j-l}}, The wheat trade networks.}\label{Plot_Resilience_DropEdges_Ascend_withLabels}
\end{figure}

After calculating the relative contributions of trade relationships to resilience, we sort the results in ascending or descending order and investigate the evolution of resilience of core or peripheral networks as trade relationships are removed cumulatively, as shown in Figs.~\ref{Plot_Resilience_DropEdges_Descend_withLabels} and \ref{Plot_Resilience_DropEdges_Ascend_withLabels}, respectively. 

\begin{table}[htb!]
\renewcommand\arraystretch{0.9}
\renewcommand{\tabcolsep}{3.5mm}
\caption{The efficiency, redundancy, and resilience of the optimal international food trade sub-network after $p$ proportion relationships are removed cumulatively in descending order of $\Delta R$.}
\label{Table_MaxResilience_Descend}
{\centering
    \begin{tabular}{lrrrrrr}
    \toprule
        ~ & 1992 & 1998 & 2004 & 2010 & 2016 & 2022  \\ \midrule
        \multicolumn{3}{l}{\it{Panel A: Maize}}    \\ 
        $p$ & 5.95\% & 5.72\% & 3.76\% & 4.48\% & 7.83\% & 4.78\%  \\ 
        Efficiency & 1.7383 & 1.8795 & 1.9145 & 1.9925 & 2.1229 & 2.0733  \\ 
        Redundancy & 2.9337 & 3.2574 & 3.2902 & 3.4206 & 3.6481 & 3.5704  \\ 
        Resilience & 0.3679 & 0.3679 & 0.3679 & 0.3679 & 0.3679 & 0.3679  \\ 
        \multicolumn{3}{l}{\it{Panel B: Rice}}  
        \\ 
        $p$ & 27.42\% & 26.32\% & 35.13\% & 40.16\% & 45.11\% & 43.37\%  \\ 
        Efficiency & 2.0214 & 2.1725 & 2.2976 & 2.3452 & 2.4174 & 2.3843  \\ 
        Redundancy & 3.4729 & 3.7278 & 3.9480 & 4.0295 & 4.1532 & 4.0953  \\ 
        Resilience & 0.3679 & 0.3679 & 0.3679 & 0.3679 & 0.3679 & 0.3679  \\ 
        \multicolumn{3}{l}{\it{Panel C: Soybean}} \\ 
        $p$ & 15.04\% & 14.15\% & 11.71\% & 8.72\% & 6.47\% & 5.86\%  \\ 
        Efficiency & 1.6021 & 1.6893 & 1.7729 & 1.8033 & 1.8719 & 1.9211  \\ 
        Redundancy & 2.7404 & 2.9093 & 3.0454 & 3.1039 & 3.2028 & 3.2823  \\ 
        Resilience & 0.3679 & 0.3679 & 0.3679 & 0.3679 & 0.3679 & 0.3679  \\ 
        \multicolumn{3}{l}{\it{Panel D: Wheat}} 
        \\         
        $p$ & 26.91\% & 29.17\% & 26.45\% & 24.25\% & 22.97\% & 21.15\%  \\ 
        Efficiency & 1.8343 & 1.9793 & 2.1144 & 2.1347 & 2.1696 & 2.1475  \\ 
        Redundancy & 3.1570 & 3.4002 & 3.6362 & 3.6681 & 3.7284 & 3.6954  \\ 
        Resilience & 0.3679 & 0.3679 & 0.3679 & 0.3679 & 0.3679 & 0.3679  \\ \bottomrule
    \end{tabular}
    }
\begin{flushleft}
\footnotesize
Notes: $p$ refers to the proportion of the number of the removed links to the total number of links.
\end{flushleft}
\end{table}

Fig.~\ref{Plot_Resilience_DropEdges_Descend_withLabels} reveals that the observed resilience enhancement directly correlates with efficiency gains, corresponding to the lack of efficiency caused by core trade relationships, aligning with previous findings \citep{FJ-Karakoc-Konar-2021-EnvironResLett}. As $p$ increases, resilience of the trade network for maize, rice, soybean, or wheat rises to reach the maximum value ($1/{\mathrm{e}}$) before declining, suggesting that there exists at least one sub-network with optimized resilience, and its metrics are shown in Table~\ref{Table_MaxResilience_Descend}. Although maize and soybean networks exhibit similar optimal $p$ values, their efficiency effects differ fundamentally. In 2022, for example, the trade-off between efficiency and redundancy in the maize trade network is due to the 68\% increase (from 1.23 to 2.07) in efficiency after core relationships are removed. However, the trade-off in the soybean trade network requires the 419\% surge (from 0.37 to 1.92) in efficiency, revealing acute oligopolistic vulnerability where the top 2 relationships contribute more than 40\% to the network resilience. Also, food trade networks exhibit sharp decline in resilience as extensive trade relationships are removed, while the phenomenon's $p$ value tends to be closer to 1 in recent years. For example, resilience of the maize trade network decreases sharply (defined as $\Delta R_p\leq-\Delta p$ at $p$) after $p=61.41\%$ trade relationships are removed in 1992, while $p=70.43\%$ and $p=91.30\%$ in 2010 and 2022, respectively (Fig.~\ref{Plot_Resilience_DropEdges_Descend_withLabels}c). This finding suggests that the capacity to sustain resilience of peripheral relationships is becoming greater and stronger.

In Fig.~\ref{Plot_Resilience_DropEdges_Ascend_withLabels}, conversely, efficiency, redundancy, and resilience remain stable until removing about 75\% peripheral trade relationships, consistently confirming that core trading relationships overwhelmingly dominate the international food trade \citep{FJ-Wood-Smith-Fanzo-Remans-DeFries-2018-NatSustain}. However, not all networks formed solely by core trade relationships attain optimal resilience, especially for soybean (Table~\ref{Table_MaxResilience_Ascend}). For instance, the 1992 soybean trade system composed of 84.96\% non-core relationships is more resilient than that composed of 14.04\% core relationships (0.3679 and 0.2260). This finding highlights the necessary roles of peripheral economies and trade relationships in sustaining international trade network resilience again, through complementary mechanisms such as improving trade efficiency and facilitating risk diversification via diversified trade pathways, indicating the irrationality of hegemonism and the importance of global cooperation.

\begin{table}[htb!]
\renewcommand\arraystretch{0.9}
\renewcommand{\tabcolsep}{3.5mm}
\caption{The efficiency, redundancy, and resilience of the optimal international food trade sub-network after $p$ proportion relationships are removed cumulatively in ascending order of $\Delta R$.}
\label{Table_MaxResilience_Ascend}
{\centering
    \begin{tabular}{lrrrrrrr}
    \toprule
        ~ & 1992 & 1998 & 2004 & 2010 & 2016 & 2022  \\ \midrule
        \multicolumn{3}{l}{\it{Panel A: Maize}}    \\ 
        $p$ & 99.52\% & 99.20\% & 98.60\% & 98.83\% & 99.53\% & 98.97\%  \\ 
        Efficiency & 0.4819 & 0.5821 & 0.7778 & 0.8902 & 0.7640 & 1.1172  \\ 
        Redundancy & 0.7326 & 1.2877 & 1.7081 & 1.5232 & 1.5169 & 1.9101  \\ 
        Resilience & 0.3668 & 0.3633 & 0.3635 & 0.3679 & 0.3664 & 0.3679  \\ 
        \multicolumn{3}{l}{\it{Panel B: Rice}}  
        \\ 
        $p$ & 94.08\% & 96.79\% & 97.95\% & 97.22\% & 99.81\% & 99.74\%  \\ 
        Efficiency & 1.3683 & 1.2987 & 1.2759 & 1.3855 & 0.6931 & 0.7627  \\ 
        Redundancy & 2.3503 & 2.2403 & 2.2038 & 2.3903 & 1.2021 & 1.3193  \\ 
        Resilience & 0.3679 & 0.3679 & 0.3679 & 0.3679 & 0.3679 & 0.3679  \\ 
        \multicolumn{3}{l}{\it{Panel C: Soybean}} \\ 
        $p$ & 85.96\% & 97.43\% & 99.08\% & 95.20\% & 69.33\% & 98.48\%  \\ 
        Efficiency & 0.3219 & 0.3551 & 0.4829 & 0.3804 & 0.3850 & 0.2883  \\ 
        Redundancy & 3.0044 & 2.2313 & 1.4116 & 2.1590 & 2.6500 & 1.7528  \\ 
        Resilience & 0.2260 & 0.2726 & 0.3484 & 0.2844 & 0.2619 & 0.2764  \\ 
        \multicolumn{3}{l}{\it{Panel D: Wheat}} 
        \\         
        $p$ & 97.67\% & 94.57\% & 96.95\% & 87.57\% & 95.26\% & 92.48\%  \\ 
        Efficiency & 0.8443 & 1.2918 & 1.2356 & 1.6963 & 1.5625 & 1.6603  \\ 
        Redundancy & 1.5266 & 2.2482 & 2.1281 & 2.9159 & 2.6974 & 2.8544  \\ 
        Resilience & 0.3677 & 0.3679 & 0.3679 & 0.3679 & 0.3679 & 0.3679  \\  \bottomrule
    \end{tabular}
    }
\begin{flushleft}
\footnotesize
Notes: $p$ refers to the proportion of the number of the removed links to the total number of links.
\end{flushleft}
\end{table}

\section{Discussions}
\label{Sec_Discussion}

\subsection{Theoretical contributions}

We have several theoretical contributions to the literature on food trade network resilience. First, our research advances the conceptual foundation of food trade system resilience by demonstrating that it can be understood as the outcome of a trade-off between efficiency and redundancy. Inspired by the assumption that long-term natural selection drives ecosystems to optimize resilience \citep{FJ-Ulanowicz-2009-EcolModel,FJ-Liang-Yu-Kharrazi-Fath-Feng-Daigger-Chen-Ma-Zhu-Mi-Yang-2020-NatFood,FJ-Kharrazi-Rovenskaya-Fath-Yarime-Kraines-2013-EcolEcon}, we employ an entropy-based approach to address the first research question. Aligning with \cite{FJ-Kharrazi-Rovenskaya-Fath-2017-PloSOne}, our analysis also suggests that systems that are either over-redundant or over-efficient are not sufficiently resilient because the former wastes excessive resources while the latter is too vulnerable during disruptions (see special cases in Appendix~\ref{Appendix_Fully} and Appendix~\ref{Appendix_Ring}). Our research not only supports the current view that resilience emerges from balancing multiple principles \citep{FJ-Essuman-Boso-Annan-2020-IntJProdEcon,FJ-Brede-deVries-2009-PhysLettA,FJ-Kamalahmadi-Shekarian-Parast-2022-IntJProdRes,FJ-Bullock-DhanjalAdams-Milne-Oliver-Todman-Whitmore-Pywell-2017-JEcol}, but also broadens the scope of resilience measurement in international food trade networks and provides an information-theoretic framework for assessing resilience across various trade networks.

Second, our research enriches resilience theory by revealing the internal mechanisms that drive resilience dynamics. By employing an index decomposition analysis \citep{FJ-Ang-Zhang-2000-Energy}, we explore how resilience evolves through the concentration of trade flows and both inter-dependence and inter-independence among economies and thereby address the second research question, in line with studies on elemental cycling networks \citep{FJ-Liang-Yu-Kharrazi-Fath-Feng-Daigger-Chen-Ma-Zhu-Mi-Yang-2020-NatFood,FJ-Luo-Yu-Kharrazi-Fath-Matsubae-Liang-Chen-Zhu-Ma-Hu-2024-NatFood}. Consistent with the existing literature \citep{FJ-Li-Wang-Kharrazi-Fath-Liu-Liu-Xiao-Lai-2024-FoodSecur}, we also examine the correlations between topological indicators and resilience. Thus, our research extends previous studies that focused mainly on external shocks or network indicators by exhibiting how resilience is shaped by internal components, thereby advancing the understanding of the structural underpinnings of resilience.

Third, our research reexamines the roles of economies and trade relationships in sustaining network resilience. Drawing on the systemic risk literature \citep{FJ-Hue-Lucotte-Tokpavi-2019-JEconDynControl, FJ-Zedda-Cannas-2020-JBankFinanc}, we define an economy or relationship as critical if its removal substantially alters the network's trade-off. While previous research has often emphasized the influence of core economies \citep{FJ-GutierrezMoya-AdensoDiaz-Lozano-2021-FoodSecur}, our findings identify that peripheral economies, through specific trade linkages, can also play important roles in maintaining resilience, which addresses the third research question. This broadens the theoretical perspective toward a more distributed and inclusive understanding of trade networks, particularly highlighting the significance of peripheral nodes. Moreover, different centrality measures, grounded in distinct assumptions \citep{FJ-Borgatti-2005-SocNetworks, FJ-Sciarra-Chiarotti-Laio-Ridolfi-2018-SciRep}, may yield divergent results. Our resilience-based perspective complements existing centrality measures and, when applied in conjunction with traditional measures, facilitates a more precise identification of nodes or links that are critical to the network structure.

\subsection{Findings and managerial implications}

The results reveal several interesting findings and inspired by these, we summarize managerial implications. First, efficiency remains the primary driver of resilience, while current international trade networks tend to exhibit greater redundancy within the $\alpha$ trade-off framework. Our findings align with views of \cite{FJ-Kharrazi-Rovenskaya-Fath-Yarime-Kraines-2013-EcolEcon}, which suggests that natural evolutionary processes are generally more effective at eliminating excess redundancy than human-designed systems. Furthermore, we find that clearer community partitions can enhance resilience by promoting denser intra-community trade flows and improving overall trade efficiency (see the analysis in \ref{Appendix_Modularity}), as observed in networks for maize, rice, and wheat. Therefore, as \cite{FJ-deRaymond-Alpha-BenAri-Daviron-Nesme-Tetart-2021-GlobFoodSecur} note, governments and regional alliances (e.g., NAFTA, MERCOSUR, G20, ASEAN, and AfCFTA) could help to maintain resilience of trade in these foods by fostering denser intra-community trade flows. For example, developing common storage facilities, harmonizing trade standards, and reducing non-tariff barriers within regional blocs would strengthen intra-community linkages, particularly during major disruptions such as the COVID-19 pandemic \citep{FJ-Wei-Xiao-Li-Huang-Liu-Xue-2025-ResourConservRecycl}. In contrast, the soybean trade network appears more vulnerable due to the imbalance between efficiency and redundancy, largely attributable to the highly monopolistic positions of major economies such as the United States, Brazil, and China, resulting in lower modularity and efficiency \citep{FJ-Li-Wang-Kharrazi-Fath-Liu-Liu-Xiao-Lai-2024-FoodSecur}. Hence, in terms of trade security and economic benefit, governments of soybean-importing economies could invest in joint regional production programs, support farmer cooperatives, and strengthen technology transfer agreements that enhance local production capacity and efficiency \citep{FJ-Ma-Zhao-Li-Niu-2024-ApplEcon}. This would not only diversify trade linkages, enhance risk resistance, and jointly ensure domestic soybean supply while promoting industrial development.

Second, rice exhibits distinct trends and characteristics compared to other staple foods, necessitating more targeted attention \citep{FJ-Burkholz-Schweitzer-2019-EnvironResLett}. We find that several topological indicators\textemdash such as network size and clustering coefficient\textemdash are positively correlated with resilience for others but not for rice. The resilience of the rice trade network tends to decline with increasing globalization, which may be attributed to its inherent edibility. However, more critical attention should be directed toward the Asia-centered monocentric structure in the rice trade and the frequent export restrictions imposed by major exporters \citep{FJ-Davis-Downs-Gephart-2021-NatFood}, both of which fundamentally undermine resilience. For example, \cite{FJ-Baum-Laber-Bruckner-Yang-Thurner-Klimek-2025-arXiv} have confirmed that Indian rice export ban substantially alter global food availability. Hence, expanding production and liberalizing rice exports are essential for improving rice trade resilience. On one hand, international organizations (e.g., FAO, WTO) should encourage diversification of rice production beyond Asia through targeted investment projects and development aid \citep{FJ-Puma-Bose-Chon-Cook-2015-EnvironResLett}. This would not only mitigate the risks of supply disruptions caused by events\textemdash such as climate extremes, exports bans, and competitive underpricing\textemdash in Asia, but also lower rice access costs in other regions and generate income for many developing countries. On the other hand, governments of major exporting countries could design export policies that allow for controlled liberalization during normal times while ensuring sufficient domestic supply. As \cite{FJ-Harold-Ashok-Valerien-Takashi-David-2024-GlobFoodSecur} note, liberalizing rice exports of major producers, while ensuring sufficient domestic supply, would benefit domestic farmers and economy.

Third, the underlying discourse on food trade has shifted from flow monopolies to strategic interactions within relational networks, placing greater emphasis on the importance of connectivity \citep{FJ-Sartori-Schiavo-2015-FoodPol}. Using index decomposition analysis, we find that the concentration effects of trade flows continue to be dominant drivers of changes in resilience, while the influence of interrelationships among economies is growing in recent years. These effects are further substantiated by the identification of critical contributors to resilience. Although our findings are broadly consistent with several studies in confirming the systemic importance of large exporters including the United States, Brazil, India, and Russia \citep{FJ-Kuhla-Kubiczek-Otto-2025-EcolEcon,FJ-Laber-Klimek-Bruckner-Yang-Thurner-2023-NatFood,FJ-Baum-Laber-Bruckner-Yang-Thurner-Klimek-2025-arXiv}, the leave-one-out analysis reveals that some peripheral economies can exert disproportionate influence on the overall resilience by sustaining irreplaceable trade pathways, which complements the above findings. For example, the existence of the link between Paraguay and Brazil improves Paraguay's indispensable role in networks for maize and rice, while Kazakhstan secures a critical position as a crisis replacement corridor. This implies that developing economies may enhance their contributions to resilience not only by increasing volumes but by adopting strategic trade positioning, such as specializing in niche commodities or maintaining bilateral agreements with key partners. For policymakers in developing countries, strengthening transport and storage infrastructure (e.g., ports, railways, grain silos) is crucial for realizing such strategies \citep{FAO-2021-State}. For donor agencies and international lenders, providing financial support for infrastructure projects directly improves resilience by enabling diversified trade routes and better market access.

Fourth, core food trade networks may undermine overall system efficiency due to the overconcentration of flows, whereas peripheral networks play a critical role in sustaining resilience by enabling risk diversification through alternative trade pathways. Within the $\alpha$ trade-off framework, we find that one of the main factors limiting efficiency in current trade systems is the overconcentration of flows along core trade links, which aligns with previous findings \citep{FJ-Karakoc-Konar-2021-EnvironResLett}, particularly for soybean. Developed economies could therefore complement their strong intra-core connections with policies that actively expand trade with developing economies, such as reducing tariffs on staple imports or supporting trade facilitation programs. This would simultaneously reduce hunger risks in vulnerable economies and diversify risks for developed ones. International organizations should play a coordinating role by promoting transparent and equitable trade rules, ensuring that peripheral actors are integrated into the global food trade system. As noted by \citet{FJ-Mujahid-Kalkuhl-2016-WorldEcon}, all economies are expected to benefit from a free and equitable food trade system that operates without discrimination.

\subsection{Limitations and future research directions}

Despite the contributions of our study, several limitations warrant further investigation. First, although the entropy-based framework provides a theoretical metric for resilience, its application should be further validated and adapted in socio-economic networks that differ from ecosystems shaped through natural selection \citep{FJ-Kharrazi-Rovenskaya-Fath-Yarime-Kraines-2013-EcolEcon}. Also, the trade-off is relative in this framework, so optimal resilience may occur even when both efficiency and redundancy are low. Therefore, the trends in efficiency and redundancy need to be jointly considered when assessing system resilience. Future research could investigate whether further autocatalytic effects emerge within the socio-economic network structure, thereby enabling redundant pathways to be replaced by more efficient alternatives. Moreover, building on different concepts and frameworks, our inferences about resilience differ from those in some studies. For example, we argue that the directed ring network is overly efficient and non-resilient (see Appendix~\ref{Appendix_Ring}), whereas \cite{FJ-Karakoc-Konar-2021-EnvironResLett} regard it as resilient. Hence, discussions toward a unified concept and framework of resilience warrant further development.

Second, we do not consider additional external drivers of resilience as other studies, such as geopolitical conflicts \citep{FJ-Laber-Klimek-Bruckner-Yang-Thurner-2023-NatFood}, extreme climate events \citep{FJ-Davis-Downs-Gephart-2021-NatFood}, and trade policy uncertainty \citep{FJ-Li-Wang-Kharrazi-Fath-Liu-Liu-Xiao-Lai-2024-FoodSecur}. Moreover, we do not apply causal inferences in this paper when investigating the interrelations between resilience and other drivers. To address these limitations, future studies could apply more advanced methods\textemdash such as machine learning\textemdash to better identify and quantify the drivers of resilience within the international food trade network, providing a more comprehensive understanding of resilience dynamics. Also, addressing these limitations could provide more robust tools for policymakers and practitioners in managing trade systems and making them more resilient.

Third, we do not account for food supply and demand dynamics when adopting the leave-one-out approach. In reality, when an economy or trade relationship is disrupted, affected economies typically adjust their trade strategies to rebalance supply and demand, triggering cascading effects that reshape flow distributions across the entire network. It is therefore important to clarify that our approach is designed solely to assess the contributions of economies or trade links to network resilience. Recent studies have employed agent-based models and stress-testing approaches to capture the dynamics of risk propagation in food systems under shock scenarios \citep{FJ-Kuhla-Kubiczek-Otto-2025-EcolEcon,FJ-Diem-Schueller-Gerschberger-Stangl-Conrady-Gerschberger-Thurner-2025-IntJProdRes}, and have further extended the analytical scope to multilayer cascading risks \citep{FJ-Naqvi-Gaupp-HochrainerStigler-2020-ORSpectrum,FJ-Laber-Klimek-Bruckner-Yang-Thurner-2023-NatFood}. Future research could build on these approaches by simulating trade flow redistributions under various shocks, which would provide deeper insights into the resilience of the international food trade system.

\section{Conclusion}
\label{Sec_Conclusion}

Based on the FAO dataset from 1986 to 2022, we construct international trade networks for maize, rice, soybean, and wheat, where economies are represented as nodes and trade relationships among them as directed and weighted links. To address the need of resilience measurement in the networked system, we employ an entropy-based approach grounded in information theory. Subsequently, we explore the correlations between resilience and topological indicators. We also employ the index decomposition analysis to quantify the relative contributions of several properties to resilience, including efficiency, redundancy, trade flows concentration, and interrelationship among economies. Furthermore, we apply the leave-one-out approach to identify critical economies and trade relationships from a resilience perspective. Given the pronounced core-periphery structure of networks, we also discuss the roles of core and peripheral components in sustaining network resilience.

Our findings provide an empirical evidence for an emerging consensus on balancing multiple principles of resilience. Given that current international trade networks are more redundant within the $\alpha$ trade-off framework, enhancing modularity and efficiency to improve resilience represents the next step in optimizing trade networks. Moreover, our findings not only suggest the dominant role of trade flow concentration, but also highlight the growing influence of interrelationships among economies, which further emphasizes the indispensable roles of some peripheral economies and trade links in sustaining resilience. From this perspective, our research highlights the attention to peripheral economies and trade relationships, while advocating for strengthened intra-regional food trade and more liberalized cross-regional food trade, thereby contributing to global food security.

\section*{Acknowledgment}

This work was partly supported by the National Natural Science Foundation of China under Grant 72171083 and the Central Universities' Program for Building World-Class Universities (Disciplines) and Special Development Guidance\textemdash Cultural Heritage and Innovation.

\section*{Disclosure statement}

No potential conflict of interest was reported by the authors.

\section*{Data availability}

Publicly available dataset is analyzed in this study. This data can be found here: \url{https://www.fao.org/faostat}.

\clearpage 

\appendix
\section*{Appendix}
\renewcommand{\thesubsection}{Appendix~\Alph{subsection}}
\renewcommand{\thesubsubsection}{\Alph{subsection}\arabic{subsubsection}}
\setcounter{subsection}{0}

\subsection{Special cases among different networks}
\label{Appendix_SpecialCases}


\subsubsection{All volume on one link}
\label{Appendix_AllVolume}

Suppose that $f_{kl} = s$ for link $\varepsilon_{kl}$ from node $k$ to node $l$ and $f_{ij}=0$ for all other links. In this case, we have $p_{kl} = 1$, $p_k^{\mathrm{out}} = 1$, $p_l^{\mathrm{in}} = 1$, and $p_{ij}=p_{i}^\mathrm{out}=p_{j}^\mathrm{in}=0$ for $\varepsilon_{ij} \neq \varepsilon_{kl}$. The joint entropy $H$ can be calculated as
\begin{equation*}
    H =
    -p_{kl}\ln p_{kl} - \sum_{\substack{(i,j) \neq (k,l)}}p_{ij}\ln p_{ij}
    =
    -\underbrace{1\cdot \ln(1)}_{\text{term }(k,l)} - \sum_{\substack{(i,j) \neq (k,l)}} \underbrace{0 \cdot \ln(\cdot)}_{=0} = 0,
\end{equation*}
while $e$ and $r$ can be calculated as
\begin{equation*}
    e = \underbrace{-1 \cdot \ln \frac{1}{1 \cdot 1}}_{\text{term } (k,l)}-\sum_{\substack{(i,j) \neq (k,l)}} \underbrace{0 \cdot \ln(\cdot)}_{=0} = 0
\end{equation*}
and
\begin{equation*}
    r = \underbrace{1 \cdot \ln \frac{1}{1}+1 \cdot \ln \frac{1}{1}}_{\text{term} (k,l)} - \sum_{\substack{(i,j) \neq (k,l)}} \underbrace{0 \cdot \ln(\cdot)}_{=0} = 0,
\end{equation*}
which indicates that the system is deterministic (i.e., has no uncertainty).


\subsubsection{No volume on some link}
\label{Appendix_NoVolume}

Suppose that link $\varepsilon_{ij}$ is structurally present, but $f_{ij}=p_{ij} \to 0$. Four situations may arise: 
\begin{enumerate}[1)]
    \item Node $i$ has no other outgoing links and node $j$ has no other incoming links such that $p_i^\mathrm{out}=p_j^\mathrm{in}=p_{ij}\to0$;
    \item Node $i$ has no other outgoing links but node $j$ has other incoming links such that $p_i^\mathrm{out}=p_{ij}\to0$ but $p_j^\mathrm{in}\neq0$; 
    \item Node $i$ has other outgoing links but node $j$ has no other incoming links such that $p_j^\mathrm{in}=p_{ij}\to0$ but $p_i^\mathrm{out}\neq0$; 
    \item Nodes $i$ has other outgoing links and node $j$ has other incoming links such that $p_i^\mathrm{out}\neq0$ and $p_j^\mathrm{in}\neq0$.
\end{enumerate}
In all of these cases, the mutual information between $i$ and $j$ can still be calculated as
\begin{equation*}
     \lim_{p_{ij} \to 0^+} p_{ij} \ln \frac{p_{ij}}{p_i^{\mathrm{out}} p_j^{\mathrm{in}}}  = 0\cdot\ln(\cdot)=0,
\end{equation*}
and the conditional entropy between $i$ and $j$ can still be calculated as
\begin{equation*}
     -\lim_{p_{ij} \to 0^+} \left(p_{ij}\ln \frac{p_{ij}}{p_{i}^{\mathrm{out}}}+ p_{ij}\ln\frac{p_{ij}}{p_{j}^{\mathrm{in}}}\right) = 0\cdot\ln(\cdot)=0,
\end{equation*}
which indicates that zero-flow pairs do not contribute to the total efficiency and redundancy.


\subsubsection{Directed fully connected trade network}
\label{Appendix_Fully}

Consider a directed fully connected trade network, i.e., a pair of reciprocal links exists between any two nodes. For simplicity, we suppose that $f_{ij}=p_{ij}=1/N_\mathcal{E}$, where $N_\mathcal{E}=N_\mathcal{V}(N_\mathcal{V}-1)$. Hence, the efficiency and redundancy can be calculated as
\begin{equation*}
    e=\sum\limits_{i=1}^{N_\mathcal{V}}\sum\limits_{j=1}^{N_\mathcal{V}} p_{ij}\ln\frac{p_{ij}}{p_{i}^{\mathrm{out}}p_{j}^{\mathrm{in}}}
    =\sum_{i=1}^{N_\mathcal{V}}\sum_{j=1}^{N_\mathcal{V}}\frac{1}{N_\mathcal{V}(N_\mathcal{V}-1)}\ln\frac{N_\mathcal{V}(N_\mathcal{V}-1)}{(N_\mathcal{V}-1)^2}
    =\ln\frac{N_\mathcal{V}}{N_\mathcal{V}-1},
\end{equation*}
and
\begin{equation*}
    r
    =-\sum\limits_{i=1}^{N_\mathcal{V}}\sum\limits_{j=1}^{N_\mathcal{V}} p_{ij}\ln \frac{p_{ij}^2}{p_{i}^{\mathrm{out}}p_{j}^{\mathrm{in}}}
    =\sum_{i=1}^{N_\mathcal{V}}\sum_{j=1}^{N_\mathcal{V}}\frac{1}{N_\mathcal{V}(N_\mathcal{V}-1)}\ln(N_\mathcal{V}-1)^2
    =2\ln(N_\mathcal{V}-1),
\end{equation*}
where $r$ exhibits a logarithmic increase but $e$ converges to $0$ if $N_\mathcal{V}$ is large.

\subsubsection{Directed ring trade network}
\label{Appendix_Ring}

Consider a directed ring trade network, i.e., each node has exactly one in-flow pathway and one out-flow pathway. For simplicity, we suppose that $f_{ij}=p_{ij}=1/N_\mathcal{E}$, where $N_\mathcal{E}=N_\mathcal{V}$. Hence, the efficiency and redundancy can be calculated as
\begin{equation*}
    e=\sum\limits_{i=1}^{N_\mathcal{V}}\sum\limits_{j=1}^{N_\mathcal{V}} p_{ij}\ln\frac{p_{ij}}{p_{i}^{\mathrm{out}}p_{j}^{\mathrm{in}}}
    =\sum_{i=1}^{N_\mathcal{V}}\sum_{j=1}^{N_\mathcal{V}}\frac{1}{N_\mathcal{V}}\ln\frac{N_\mathcal{V}^2}{N_\mathcal{V}}
    =\ln N_\mathcal{V},
\end{equation*}
and
\begin{equation*}
    r
    =-\sum\limits_{i=1}^{N_\mathcal{V}}\sum\limits_{j=1}^{N_\mathcal{V}} p_{ij}\ln \frac{p_{ij}^2}{p_{i}^{\mathrm{out}}p_{j}^{\mathrm{in}}}
    =\sum_{i=1}^{N_\mathcal{V}}\sum_{j=1}^{N_\mathcal{V}}\frac{1}{N_\mathcal{V}}\ln\frac{N_\mathcal{V}^2}{N_\mathcal{V}^2}
    =0,
\end{equation*}
where $e$ exhibits a logarithmic increase if $N_\mathcal{V}$ is large. It also indicates that the system is deterministic (i.e., has no uncertainty).

These cases confirm that the $\alpha$ resilience framework is well-defined for non-negative flows ($f_{ij} \geq 0$ and $p_{ij} \geq 0$), and behaves consistently in scenarios such as deterministic cases, zero-flow systems, fully connected networks, and ring networks.

\subsection{Modularity and resilience}
\label{Appendix_Modularity}

Modularity can be expressed as:
\begin{equation*}
    Q = \frac{1}{N_\mathcal{V}\langle s\rangle} \sum\limits_{i=1}^{N_\mathcal{V}}\sum\limits_{j=1}^{N_\mathcal{V}}\left(f_{ij}-\frac{s_is_j}{N_\mathcal{V}\langle s\rangle}\right)\delta(c_i,c_j)
    =
    \frac{1}{s} \sum\limits_{i=1}^{N_\mathcal{V}}\sum\limits_{j=1}^{N_\mathcal{V}}\left(f_{ij}-\frac{s_is_j}{s}\right)\delta(c_i,c_j),
\end{equation*}
where $s$ denotes the the total system flow. Relatively large modularity values between 0.3 and 0.7 are considered an indicator of a well-structured network partition \citep{FJ-Newman-2004-PhysRevE}, which represents that the actual flows are higher than the expected flows within communities, expressed as:
\begin{equation*}
    f_{i,j}\gg \frac{s_is_j}{s},
\end{equation*}
which denotes that flows within communities are inter-dependent, corresponding to the higher efficiency in Eq.~(\ref{eq_efficiency}) and pmi in Eq.~(\ref{eq_pmi}). Since the current networks are more redundant within the $\alpha$ framework, the higher efficiency is, the higher resilience is. Hence, for the current international food trade networks with stronger community structures, $Q\propto e\propto R$.

\subsection{Community partition for food trade networks}
\label{Appendix_Community}

In this paper, we use the Louvain algorithm for community detection and examine the evolution of community structures across three periods (Fig.~\ref{Plot_Map_Community_withLabels}). Economies within the same community are typically clustered in contiguous geographic regions, a pattern that can be largely explained by transport costs, regional trade policies, and geopolitical ties. For example, European countries consistently form a single community; North Africa and the Middle East often constitute a distinct bloc or alternatively maintain stable linkages with Europe; and South American countries largely cluster together. Geographic proximity fosters closer trade linkages among neighboring economies, thereby facilitating risk distribution at lower transportation costs in the event of localized shocks and ultimately enhancing overall resilience.

Beyond geography, food-specific attributes also shape community evolution. For soybean, the persistent dominance of the United States, Brazil, and Argentina as major producers and exporters has fostered inter-community connectivity but simultaneously constrained the emergence of diversified community structures. This concentration weakens modular boundaries and reduces the contribution of community diversity to resilience. By contrast, rice and wheat production and exports are more geographically dispersed, resulting in more distinct and diverse community structures, thereby sustaining resilience by strengthening community-based risk buffering.

\renewcommand{\thefigure}{\Alph{subsection}\arabic{figure}}
\begin{figure}[htb!]
\centering 
\includegraphics[width=1\textwidth]{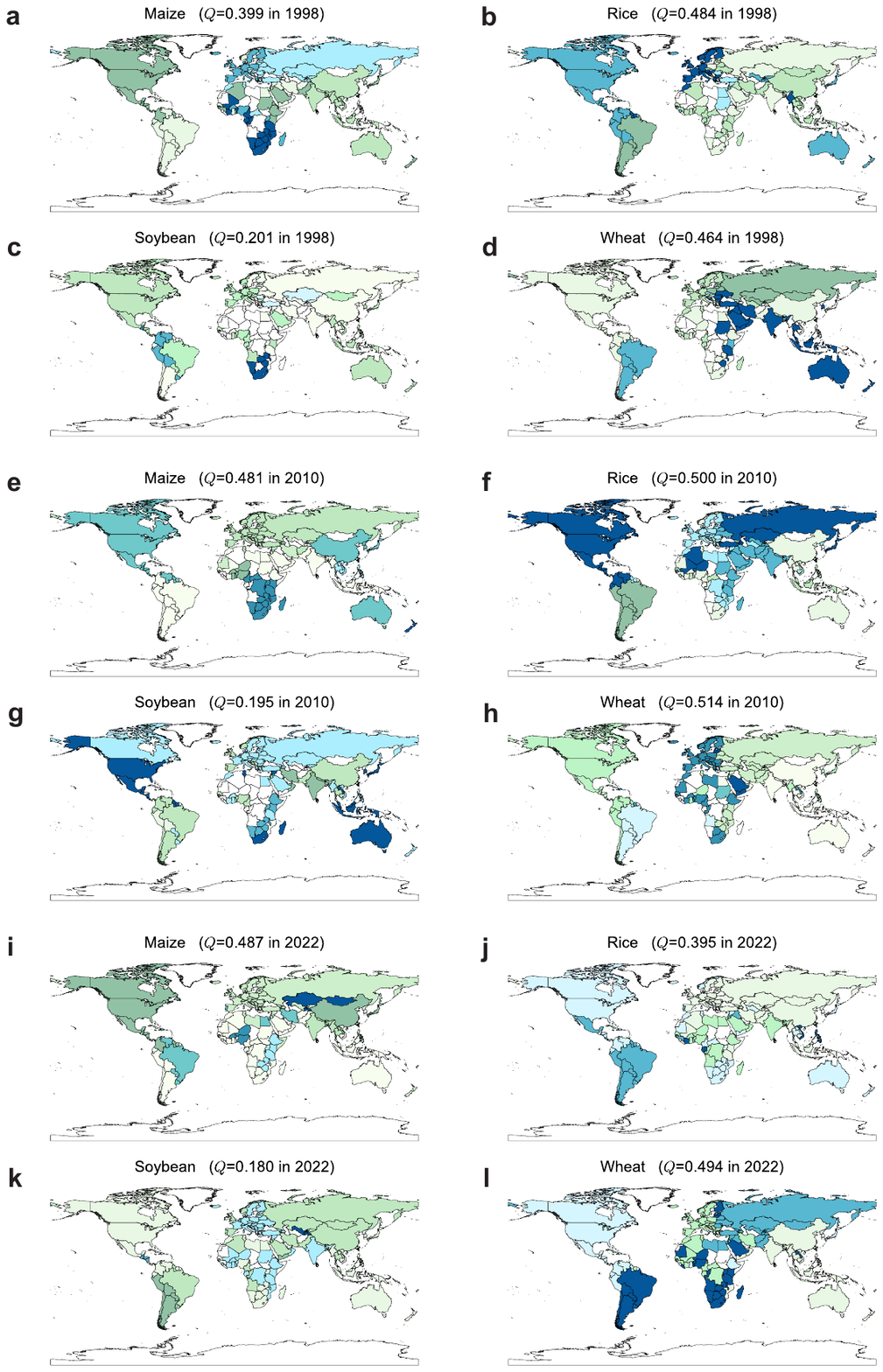} 
\caption{Community structures in the international trade networks for maize, rice, soybean, and wheat. In this figure, a blank indicates that: 1) no economy here, 2) the economy did not participate in trade, or 3) data are missing. {\bf{a-d}}, In 1998. {\bf{c-h}}, In 2010. {\bf{i-l}}, In 2022. }
\label{Plot_Map_Community_withLabels}
\end{figure}

\clearpage

\bibliographystyle{elsarticle-harv}
\bibliography{Bib1,Bib2,BibRCE,BibITN,BibRobustNet}

\end{document}